\documentclass[12pt,a4paper]{article}
\usepackage{tikz}
\usepackage{pgf}

\usepackage{amsmath}
\usepackage{booktabs}
\usepackage{cite}
\usepackage[colorlinks=true,linkcolor=black, citecolor=black,urlcolor=black]{hyperref}

\usepackage{multirow,graphics}
\usepackage{enumerate}
\usepackage{amstext}
\usepackage{amssymb}
\usepackage{amsmath}

\usepackage{url}
\usepackage{graphicx}
\usepackage{subcaption}
\usepackage{color}
\usepackage[nodisplayskipstretch]{setspace}
\usepackage{tcolorbox}
\usepackage{float}
\usepackage{wasysym}
\usepackage{ytableau}
\usepackage{textcomp}

\usetikzlibrary{decorations.pathmorphing}
\usetikzlibrary{decorations.markings}
\usetikzlibrary{intersections}
\usetikzlibrary{calc}
\usetikzlibrary{positioning}
\usetikzlibrary{3d}
\usetikzlibrary{shapes}
\usetikzlibrary{angles}
\usetikzlibrary{fit}
\usetikzlibrary{backgrounds}

\tikzset{
photon/.style={decorate, decoration={snake}},
particle/.style={postaction={decorate},
    decoration={markings,mark=at position .5 with {\arrow{>}}}},
antiparticle/.style={postaction={decorate},
    decoration={markings,mark=at position .5 with {\arrow{<}}}},
gluon/.style={decorate, decoration={coil,amplitude=2pt, segment length=4pt},color=purple},
wilson/.style={color=blue, thick},
scalarZ/.style={postaction={decorate},decoration={markings, mark=at position .5 with{\arrow[scale=1]{stealth}}}},
scalarX/.style={postaction={decorate}, dashed, dash pattern = on 4pt off 2pt, dash phase = 2pt, decoration={markings, mark=at position .53 with{\arrow[scale=1]{stealth}}}},
scalarZw/.style={postaction={decorate},decoration={markings, mark=at position .75 with{\arrow[scale=1]{stealth}}}},
scalarXw/.style={postaction={decorate}, dashed, dash pattern = on 4pt off 2pt, dash phase = 2pt, decoration={markings, mark=at position .60 with{\arrow[scale=1]{stealth}}}},
frozen/.style={inner sep=0.7mm, rectangle,draw},
frozenblue/.style={rectangle, draw, fill=blue!20, inner sep=0.7mm},
norm/.style={->, draw, shorten <=2pt, shorten >=2pt},
diag/.style={->, draw, shorten <=5pt, shorten >=3pt},
every node/.style={inner sep=0.5mm},
webarrow/.style={postaction={decorate},,decoration={markings, mark=at position .5 with{\arrow[scale=1]{stealth}}}}
}

\newcommand{\as}[2]{
  $a_{#1#2}$
}

\makeatletter
\tikzoption{canvas is plane}[]{\@setOxy#1}
\def\@setOxy O(#1,#2,#3)x(#4,#5,#6)y(#7,#8,#9)%
  {\def\tikz@plane@origin{\pgfpointxyz{#1}{#2}{#3}}%
   \def\tikz@plane@x{\pgfpointxyz{#4}{#5}{#6}}%
   \def\tikz@plane@y{\pgfpointxyz{#7}{#8}{#9}}%
   \tikz@canvas@is@plane
  }
\makeatother

\newcommand\be{\begin{equation}}
\newcommand\ee{\end{equation}}

\newcommand{\cA}{\mathcal{A}}

\normalsize
\makeatletter
\divide\@tempdima\baselineskip
\@tempcnta=\@tempdima
\skip\@mpfootins = \skip\footins
\renewcommand{\@dotsep}{10000}
\makeatother

\begin{document}
\numberwithin{equation}{section}
\begin{center}
\phantom{vv}

\vspace{3cm}
\bigskip

{\Large \bf Tropical fans, scattering equations and amplitudes}

\bigskip
\mbox{\bf James Drummond\footnotemark, Jack Foster$^1$,  \"Omer G\"urdo\u gan\footnotemark, Chrysostomos Kalousios$^1$}%
\setcounter{footnote}{1}
\footnotetext{\, {\texttt{\{j.a.foster, j.m.drummond, c.kalousios\}@soton.ac.uk }}}
\setcounter{footnote}{2}
\footnotetext{\,\,\texttt{omer.gurdogan@maths.ox.ac.uk}}
\bigskip

{$^1$\em School of Physics \& Astronomy, University of Southampton,\\
  Highfield, Southampton, SO17 1BJ, United Kingdom.}\\[10pt]

{$^2$\em Mathematical Institute, University of Oxford,\\
Andrew Wiles Building,
Woodstock Road,
Oxford,
OX2 6GG, United Kingdom.}

\vspace{3cm}  {\bf Abstract}
\end{center}

We describe a family of tropical fans related to Grassmannian cluster algebras. These fans are related to the kinematic space of massless scattering processes in a number of ways. For each fan associated to the Grassmannian ${\rm Gr}(k,n)$ there is a notion of a generalised $\phi^3$ amplitude and an associated set of scattering equations which further generalise the ${\rm Gr}(k,n)$ scattering equations that have been recently introduced. Here we focus mostly on the cases related to finite Grassmannian cluster algebras and we explain how face variables for the cluster polytopes are simply related to the scattering equations. For the Grassmannians ${\rm Gr}(4,n)$ the tropical fans we describe are related to the singularities (or symbol letters) of loop amplitudes in planar $\mathcal{N}=4$ super Yang-Mills theory. We show how each choice of tropical fan leads to a natural class of polylogarithms, generalising the notion of cluster adjacency and we describe how the currently known loop data fit into this classification.

\noindent

\newpage
\phantom{vv}
\vspace{1cm}
\hrule
\tableofcontents

\bigskip
\medskip

\hrule

\section{Introduction}

Recently many connections have been made between the study of tropical geometry and scattering amplitudes in quantum field theory and string theory. One connection is via the study of massless scattering amplitudes via the scattering equations \cite{Cachazo:2013gna,Cachazo:2013hca,Cachazo:2013iea}. In the simplest setting these equations describe tree-level biadjoint $\phi^3$ amplitudes. In this context there is an auxiliary space, the moduli space of $n$ points on a Riemann sphere, which is related to the kinematics of the $n$-point massless scattering amplitude via the scattering equations. The moduli space is the configuration space of $n$ points in $\mathbb{P}^1$ which is equivalent to the Grassmannian ${\rm Gr}(2,n)$ (modulo local rescalings). The biadjoint $\phi^3$ amplitude can be computed by evaluating certain Parke-Taylor type factors on solutions of the scattering equations. The final expression obtained after summing all solutions is equal to the traditional Feynman diagram expression. 

In \cite{Cachazo:2019ngv}, a connection of the above picture to the tropical Grassmannian was made. The tropical version of a space is a simplification in which the defining non-linear equations are treated in a piecewise linear fashion. Despite this simplification, the tropical space retains much information from the original. In particular, each $\phi^3$ Feynman diagram can be associated to a maximal cone of the tropical Grassmannian ${\rm Gr}(2,n)$. This picture is closely related \cite{Drummond:2019qjk} to the kinematic associahredron picture of \cite{Arkani-Hamed:2017mur}.

Moreover, in \cite{Cachazo:2019ngv} a generalisation of the above picture to
Grassmannians ${\rm Gr}(k,n)$ was given. In the generalised setting
of ${\rm Gr}(k,n)$ there is no known standard field theory formulation for the amplitudes.
However it was conjectured that the associated generalised amplitude obtained from the scattering equations can again be written as a sum over maximal cones of the tropical Grassmannian. 

Another setting in which the Grassmannian space arises is in the study of loop amplitudes in planar $\mathcal{N}=4$ super Yang-Mills theory. In this case the kinematic space itself can be identified with the Grassmannian ${\rm Gr}(4,n)$ via the introduction of momentum twistor variables \cite{Hodges:2009hk}. As described in \cite{Golden:2013xva}, there is a close connection between the branch cut singularities of the loop amplitudes and the cluster algebras of \cite{1021.16017,1054.17024,fomin_zelevinsky_2007} as applied to Grassmannian spaces \cite{scott_2006}. The Grassmannian cluster algebra is itself closely connected to the tropical Grassmannian, or more specifically, its positive part. 

Tropical Grassmannians were initially studied in \cite{SpeyerSturmfels} and the positive part described in \cite{2003math.....12297S}. In \cite{Drummond:2019qjk} we described how the technology of cluster algebras, in particular the idea of mutations and ${\bf g}$-vector fans can be a useful tool in the study of tropical Grassmannians and hence the generalised $\phi^3$ amplitudes.
Here we will study further the connection between cluster algebras and the positive tropical Grassmannian, a link already partly explored in \cite{2003math.....12297S}. We will identify a whole range of tropical fans which can be associated with the positive tropical Grassmannian, one of which is the fan of \cite{2003math.....12297S} and another is the ${\bf g}$-vector fan of the cluster algebra. In the finite cases the ${\bf g}$-vector fan (which we refer to as the `cluster' fan) is the most refined fan and other fans we consider, including the fan of \cite{2003math.....12297S}, can be obtained as projections of it.

These considerations lead us to propose new scattering equations which are more general than those of \cite{Cachazo:2019ngv} and involve a more general set of Mandelstam invariants. We can obtain generalised amplitudes which depend on the generalised Mandelstam invariants in a similar fashion by considering volumes of facets of the corresponding tropical fan. This construction can also be used to describe the dual cluster polytope by providing a direct route to determining the face variables, which define the codimension-one boundaries of these polytopes. In this regard the tropical fans we study and their associated generalised $\phi^3$ ampliutudes are very closely related to the notion of stringy canonical forms introduced in \cite{Arkani-Hamed:2019mrd,Herderschee:2019wtl,Arkani-Hamed:2019plo}. Indeed the integrals considered there provide in principle a deformation of the $\phi^3$ amplitude in the same way that tree-level superstring amplitudes are effectively derived from the $\alpha'$ deformation of biadjoint $\phi^3$ amplitudes. In fact a range of techniques explored in recent papers are effectively different languages to describe the same (or closely related) underlying mathematics, namely the tropical Grassmannians discussed in \cite{Cachazo:2019ngv}, the cluster algebras, mutations and ${\bf g}$-vector fans as studied in \cite{Drummond:2019qjk,Drummond:2019cxm,Henke:2019hve}, the Minkowski sums of Newton polytopes \cite{Arkani-Hamed:2019rds,Arkani-Hamed:2019mrd,He:2020ray} (which are dual to tropical fans), the planar arrangements of \cite{Borges:2019csl,Cachazo:2019xjx} and matroid subdivisions, as studied in \cite{Early:2019zyi,Early:2019eun}.

With a selection of different tropical fans to hand we discuss how such differences may show up in the singularities of loop amplitudes in $\mathcal{N}=4$ super Yang-Mills theory. This leads us to a generalisation of the notion of `cluster adjacency' put forward in \cite{Drummond:2017ssj}. Here we define a natural set of polylogarithms satisfying adjacency criteria (not only pairs but also triplets, and in general longer consecutive sequences). The cluster adjacent polylogarithms of \cite{Drummond:2017ssj} correspond to the cluster fan, while less refined fans lead to stronger sets of adjacency criteria. We examine the known loop amplitudes and compare them at the level of pairs and triplets to establish a tentative correspondence between amplitudes of different MHV degree and classes of tropical fans.

\section{Grassmannian cluster algebras and tropical fans}
\label{Sect-fans}

Let us begin by recalling the construction of Speyer and Williams \cite{2003math.....12297S} to describe the positive part of the tropical Grassmannian ${\rm Gr}(k,n)$. We will focus on ${\rm Gr}(3,6)$ as our motivating example and later consider also ${\rm Gr}(3,7)$ and ${\rm Gr}(3,8)$. Apart from the Grassmannians ${\rm Gr}(2,n)$ these cases (and their duals) exhaust the list of finite cluster Grassmannian cluster algebras. Of particular relevance to planar amplitudes in $\mathcal{N}=4$ super Yang-Mills theory is the case ${\rm Gr}(3,7)$ which is dual to ${\rm Gr}(4,7)$. We also provide some more details on the infinite case ${\rm Gr}(4,8)$ studied in \cite{Drummond:2019cxm}.

\begin{figure}
\begin{center}
{\footnotesize
\begin{tikzpicture}
\pgfmathsetmacro{\nw}{1.1}
\pgfmathsetmacro{\vvwnw}{1.9}
\pgfmathsetmacro{\vvvwnw}{2.25}
\pgfmathsetmacro{\nh}{0.6}
\pgfmathsetmacro{\aa}{0.6}
\pgfmathsetmacro{\ep}{0.1}
\node at (-0.5*\nw -\aa,\aa+0.5*\nh) {$\langle 1\,2\,3 \rangle$};
\draw[] (-\aa,\aa) -- (-\aa -\nw,\aa) -- (-\aa -\nw, \aa+\nh) -- (-\aa,\aa+\nh) -- cycle;
\node at (0.5*\nw +0*\aa,-0*\aa-0.5*\nh) {$\langle 1\,2\,4 \rangle$};
\node at (0.5*\nw +0*\aa,-1*\aa-1.5*\nh) {$\langle 1\,3\,4 \rangle$};
\node at (0.5*\nw +0*\aa,-2*\aa-2.5*\nh) {$\langle 2\,3\,4 \rangle$};
\draw[] (0,-2*\aa-2*\nh) -- (0,-2*\aa-3*\nh) -- (\nw,-2*\aa-3*\nh) -- (\nw,-2*\aa-2*\nh) -- cycle;
\node at (1.5*\nw +1*\aa,-0*\aa-0.5*\nh) {$\langle 1\,2\,5 \rangle$};
\node at (1.5*\nw +1*\aa,-1*\aa-1.5*\nh) {$\langle 1\,4\,5 \rangle$};
\node at (1.5*\nw +1*\aa,-2*\aa-2.5*\nh) {$\langle 3\,4\,5 \rangle$};
\draw[] (\nw+\aa,-2*\aa-2*\nh) -- (\nw+\aa,-2*\aa-3*\nh) -- (2*\nw+\aa,-2*\aa-3*\nh) -- (2*\nw+\aa,-2*\aa-2*\nh) -- cycle;
\node at (3*\nw + 2*\aa+0.5*\vvvwnw,-0*\aa-0.5*\nh) {$\langle 1\,2\,n \scalebox{0.65}[1.0]{\( - \)} 1 \rangle$};
\node at (3*\nw + 2*\aa+0.5*\vvvwnw,-1*\aa-1.5*\nh) {$\langle 1\,n \scalebox{0.65}[1.0]{\( - \)} 2\,n \scalebox{0.65}[1.0]{\( - \)} 1 \rangle$};
\node at (3*\nw + 2*\aa+0.5*\vvvwnw,-2*\aa-2.5*\nh) {$\langle n \scalebox{0.65}[1.0]{\( - \)} 3\,n \scalebox{0.65}[1.0]{\( - \)} 2\,n \scalebox{0.65}[1.0]{\( - \)} 1 \rangle$};
\draw[] (3*\nw+2*\aa,-2*\aa-2*\nh) -- (3*\nw+2*\aa,-2*\aa-3*\nh) -- (3*\nw+2*\aa+\vvvwnw,-2*\aa-3*\nh) -- (3*\nw+2*\aa+\vvvwnw,-2*\aa-2*\nh) -- cycle;
\node at (3*\nw + 3*\aa+1*\vvvwnw+0.5*\vvwnw,-0*\aa-0.5*\nh) {$\langle 1\,2\,n \rangle$};
\draw[] (3*\nw+3*\aa+1*\vvvwnw,-0*\aa-0*\nh) -- (3*\nw+3*\aa+1*\vvvwnw,-0*\aa-1*\nh) -- (3*\nw+3*\aa+1*\vvvwnw+1*\vvwnw,-0*\aa-1*\nh) -- (3*\nw+3*\aa+1*\vvvwnw+1*\vvwnw,-0*\aa-0*\nh) -- cycle;
\node at (3*\nw + 3*\aa+1*\vvvwnw+0.5*\vvwnw,-1*\aa-1.5*\nh) {$\langle 1\,n \scalebox{0.65}[1.0]{\( - \)} 1\,n \rangle$};
\draw[] (3*\nw+3*\aa+1*\vvvwnw,-1*\aa-1*\nh) -- (3*\nw+3*\aa+1*\vvvwnw,-1*\aa-2*\nh) -- (3*\nw+3*\aa+1*\vvvwnw+1*\vvwnw,-1*\aa-2*\nh) -- (3*\nw+3*\aa+1*\vvvwnw+1*\vvwnw,-1*\aa-1*\nh) -- cycle;
\node at (3*\nw + 3*\aa+1*\vvvwnw+0.5*\vvwnw,-2*\aa-2.5*\nh) {$\langle n \scalebox{0.65}[1.0]{\( - \)} 2\,n \scalebox{0.65}[1.0]{\( - \)} 1\,n \rangle$};
\draw[] (3*\nw+3*\aa+1*\vvvwnw,-2*\aa-2*\nh) -- (3*\nw+3*\aa+1*\vvvwnw,-2*\aa-3*\nh) -- (3*\nw+3*\aa+1*\vvvwnw+1*\vvwnw,-2*\aa-3*\nh) -- (3*\nw+3*\aa+1*\vvvwnw+1*\vvwnw,-2*\aa-2*\nh) -- cycle;
\node at (2.25*\nw + 2*\aa,-0.5*\nh) {$\ldots$};
\node at (2.25*\nw + 2*\aa,-1.5*\nh-\aa) {$\ldots$};
\node at (2.25*\nw + 2*\aa,-2.5*\nh-2*\aa) {$\ldots$};
\draw[->] (-\aa+0*\ep,\aa-\ep) -- (0-0*\ep,0+\ep);
\draw[->] (0.5*\nw,-\nh-\ep) -- (0.5*\nw,-\nh-\aa+\ep);
\draw[->] (0.5*\nw,-2*\nh-\aa-\ep) -- (0.5*\nw,-2*\nh-2*\aa+\ep);
\draw[->] (1*\nw+\ep,-0.5*\nh) -- (1*\nw+\aa-\ep,-0.5*\nh);
\draw[->] (1*\nw+\ep,-1.5*\nh-\aa) -- (1*\nw+\aa-\ep,-1.5*\nh-\aa);
\draw[->] (1*\nw+\aa-0*\ep,-\nh-\aa+\ep) -- (1*\nw+0*\ep,-\nh-\ep);
\draw[->] (1*\nw+\aa-0*\ep,-2*\nh-2*\aa+\ep) -- (1*\nw+0*\ep,-2*\nh-1*\aa-\ep);
\draw[->] (1.5*\nw+1*\aa,-\nh-\ep) -- (1.5*\nw+1*\aa,-\nh-\aa+\ep);
\draw[->] (1.5*\nw+1*\aa,-2*\nh-1*\aa-\ep) -- (1.5*\nw+1*\aa,-2*\nh-2*\aa+\ep);
\draw[->] (3*\nw+0.5*\vvvwnw+2*\aa,-\nh-\ep) -- (3*\nw+0.5*\vvvwnw+2*\aa,-\nh-\aa+\ep);
\draw[->] (3*\nw+0.5*\vvvwnw+2*\aa,-2*\nh-\aa-\ep) -- (3*\nw+0.5*\vvvwnw+2*\aa,-2*\nh-2*\aa+\ep);
\draw[->] (3*\nw+2*\aa+1*\vvvwnw+\ep,-0.5*\nh) -- (3*\nw+3*\aa+1*\vvvwnw-\ep,-0.5*\nh);
\draw[->] (3*\nw+2*\aa+1*\vvvwnw+\ep,-1.5*\nh-1*\aa) -- (3*\nw+3*\aa+1*\vvvwnw-\ep,-1.5*\nh-1*\aa);
\draw[->] (3*\nw+3*\aa+1*\vvvwnw-0*\ep,-\nh-\aa+\ep) -- (3*\nw+2*\aa+1*\vvvwnw+0*\ep,-\nh-\ep);
\draw[->] (3*\nw+3*\aa+1*\vvvwnw-0*\ep,-2*\nh-2*\aa+\ep) -- (3*\nw+2*\aa+1*\vvvwnw+0*\ep,-2*\nh-1*\aa-\ep);
\end{tikzpicture}
}
\end{center}
\caption{\small The initial cluster of the Grassmannian cluster algebra ${\rm Gr}(3,n)$.}
\label{Gr3ninitial}
\end{figure}
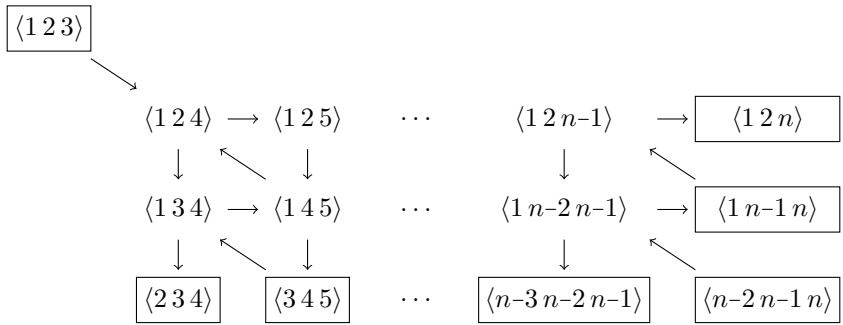

We recall the structure of the initial cluster of the Grassmannian cluster algebra ${\rm Gr}(k,n)$. The example of ${\rm Gr}(3,n)$ is shown in Fig. \ref{Gr3ninitial}. In general, the active nodes form a $(k-1)\times (n-k-1)$ array and there are also $k$ frozen nodes (depicted in boxes in Fig. \ref{Gr3ninitial}). From the initial cluster we obtain a $(k-1)\times (n-k-1)$ array of cluster $\mathcal{X}$-coordinates $x_{rs}$ given by the product of incoming $\mathcal{A}$-coordinates over the product of outgoing ones to the node in row $r$ and column $s$.

Given the $\mathcal{X}$-coordinates we can form the $(k \times n)$ web matrix
\be
W = ( 1 \!\! 1_k | M)\,,
\ee
where $M$ is the $k \times (n-k)$ matrix with entries
\be
m_{ij} = (-1)^{i+k} \sum_{\underline{\lambda}\in Y_{ij}} \prod_{r=1}^{k-i} \prod_{s=1}^{\lambda_r} x_{rs}\,.
\label{mentries}
\ee
The sum in (\ref{mentries}) is over the range $Y_{ij}$ given by $0\leq \lambda_{k-i} \leq \ldots \leq \lambda_1 \leq j-1$.
We can then evaluate all the $\mathcal{A}$-coordinates as polynomials the $\mathcal{X}$-coordinates with positive coefficients by identifying the Pl\"ucker coordinates $\langle i_1 \ldots i_k \rangle$ with the maximal minors formed by taking the columns $i_1,\ldots,i_k$ of the web matrix $W$.

\begin{figure}
\begin{center}
{\footnotesize
\begin{tikzpicture}
\pgfmathsetmacro{\nw}{1.3}
\pgfmathsetmacro{\vvwnw}{2.5}
\pgfmathsetmacro{\vvvwnw}{2.85}
\pgfmathsetmacro{\nh}{0.6}
\pgfmathsetmacro{\aa}{0.6}
\pgfmathsetmacro{\ep}{0.1}
\node at (-0.5*\nw -\aa,\aa+0.5*\nh) {$\langle 1\,2\,3 \rangle$};
\draw[] (-\aa,\aa) -- (-\aa -\nw,\aa) -- (-\aa -\nw, \aa+\nh) -- (-\aa,\aa+\nh) -- cycle;
\node at (0.5*\nw +0*\aa,-0*\aa-0.5*\nh) {$\langle 1\,2\,4 \rangle$};
\node at (0.5*\nw +0*\aa,-1*\aa-1.5*\nh) {$\langle 1\,3\,4 \rangle$};
\node at (0.5*\nw +0*\aa,-2*\aa-2.5*\nh) {$\langle 2\,3\,4 \rangle$};
\draw[] (0,-2*\aa-2*\nh) -- (0,-2*\aa-3*\nh) -- (\nw,-2*\aa-3*\nh) -- (\nw,-2*\aa-2*\nh) -- cycle;
\node at (1.5*\nw +1*\aa,-0*\aa-0.5*\nh) {$\langle 1\,2\,5 \rangle$};
\node at (1.5*\nw +1*\aa,-1*\aa-1.5*\nh) {$\langle 1\,4\,5 \rangle$};
\node at (1.5*\nw +1*\aa,-2*\aa-2.5*\nh) {$\langle 3\,4\,5 \rangle$};
\draw[] (\nw+\aa,-2*\aa-2*\nh) -- (\nw+\aa,-2*\aa-3*\nh) -- (2*\nw+\aa,-2*\aa-3*\nh) -- (2*\nw+\aa,-2*\aa-2*\nh) -- cycle;
\node at (2.5*\nw +2*\aa,-0*\aa-0.5*\nh) {$\langle 1\,2\,6 \rangle$};
\draw[] (2*\nw+2*\aa,-0*\aa-0*\nh) -- (2*\nw+2*\aa,-0*\aa-1*\nh) -- (3*\nw+2*\aa,-0*\aa-1*\nh) -- (3*\nw+2*\aa,-0*\aa-0*\nh) -- cycle;
\node at (2.5*\nw +2*\aa,-1*\aa-1.5*\nh) {$\langle 1\,5\,6 \rangle$};
\draw[] (2*\nw+2*\aa,-1*\aa-1*\nh) -- (2*\nw+2*\aa,-1*\aa-2*\nh) -- (3*\nw+2*\aa,-1*\aa-2*\nh) -- (3*\nw+2*\aa,-1*\aa-1*\nh) -- cycle;
\node at (2.5*\nw +2*\aa,-2*\aa-2.5*\nh) {$\langle 4\,5\,6 \rangle$};
\draw[] (2*\nw+2*\aa,-2*\aa-2*\nh) -- (2*\nw+2*\aa,-2*\aa-3*\nh) -- (3*\nw+2*\aa,-2*\aa-3*\nh) -- (3*\nw+2*\aa,-2*\aa-2*\nh) -- cycle;
\draw[->] (-\aa+0*\ep,\aa-\ep) -- (0-0*\ep,0+\ep);
\draw[->] (0.5*\nw,-\nh-\ep) -- (0.5*\nw,-\nh-\aa+\ep);
\draw[->] (0.5*\nw,-2*\nh-\aa-\ep) -- (0.5*\nw,-2*\nh-2*\aa+\ep);
\draw[->] (1*\nw+\ep,-0.5*\nh) -- (1*\nw+\aa-\ep,-0.5*\nh);
\draw[->] (1*\nw+\ep,-1.5*\nh-\aa) -- (1*\nw+\aa-\ep,-1.5*\nh-\aa);
\draw[->] (1*\nw+\aa-0*\ep,-\nh-\aa+\ep) -- (1*\nw+0*\ep,-\nh-\ep);
\draw[->] (1*\nw+\aa-0*\ep,-2*\nh-2*\aa+\ep) -- (1*\nw+0*\ep,-2*\nh-1*\aa-\ep);
\draw[->] (1.5*\nw+1*\aa,-\nh-\ep) -- (1.5*\nw+1*\aa,-\nh-\aa+\ep);
\draw[->] (1.5*\nw+1*\aa,-2*\nh-1*\aa-\ep) -- (1.5*\nw+1*\aa,-2*\nh-2*\aa+\ep);
\draw[->] (2*\nw+\aa+\ep,-0.5*\nh) -- (2*\nw+2*\aa-\ep,-0.5*\nh);
\draw[->] (2*\nw+\aa+\ep,-1.5*\nh-\aa) -- (2*\nw+2*\aa-\ep,-1.5*\nh-\aa);
\draw[->] (2*\nw+2*\aa-0*\ep,-\nh-\aa+\ep) -- (2*\nw+\aa+0*\ep,-\nh-\ep);
\draw[->] (2*\nw+2*\aa-0*\ep,-2*\nh-2*\aa+\ep) -- (2*\nw+\aa+0*\ep,-2*\nh-1*\aa-\ep);
\end{tikzpicture}
}
\end{center}
\caption{\small The initial cluster of the Grassmannian cluster algebra ${\rm Gr}(3,6)$. 
}
\label{Gr36initial}
\end{figure}
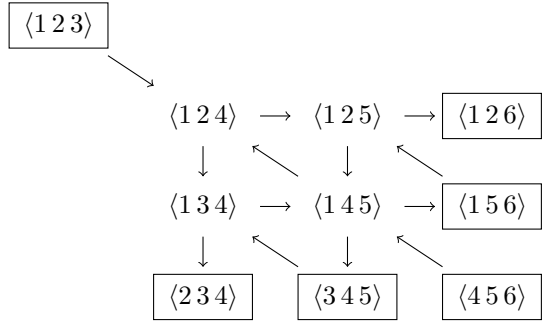

In the case of ${\rm Gr}(3,6)$ we have the initial cluster shown in Fig \ref{Gr36initial}. For this cluster we have the following cluster $\mathcal{X}$-coordinates,
\begin{align}
x_{11} &= \frac{\langle 123 \rangle \langle 145 \rangle }{\langle 125 \rangle \langle 134 \rangle}\,, \quad &&x_{12} = \frac{\langle 124 \rangle \langle 156 \rangle}{\langle 126 \rangle \langle 145 \rangle} \,, \notag \\
x_{21} &= \frac{\langle 124 \rangle \langle 345 \rangle }{\langle 234 \rangle \langle 145 \rangle}\,, \quad &&x_{22} = \frac{\langle 134 \rangle \langle 456 \rangle \langle 125 \rangle}{\langle 124 \rangle \langle 345 \rangle \langle 156 \rangle}\,.
\end{align}
The web matrix then takes the form $W= (1\!\!1_3 | M )$ with
\be
{\small
M\!=\!\left[
\begin{matrix}
1 &\, & 1 + x_{11} + x_{11} x_{21} &\, & 1+ x_{11} + x_{11} x_{21} + x_{11} x_{12} + x_{11} x_{12} x_{21} + x_{11} x_{12} x_{21} x_{22}\\
-1 &\, & -1 - x_{11} &\, &-1-x_{11} - x_{11} x_{12}\\
1 & \,& 1 &\, & 1
\end{matrix}
\right]
}
\label{webmatrix}
\ee
If we identify the Pl\"ucker coordinate $\langle ijk \rangle$ with the minor formed by taking columns $i$, $j$ and $k$ of the web matrix then we find that all the $\mathcal{A}$ coordinates of the ${\rm Gr}(3,6)$ cluster algebra are expressed as polynomials in the $\mathcal{X}$-coordinates. To emphasise this point we also use the notation $p_{ijk} = \langle ijk \rangle$. The frozen $\mathcal{A}$-coordinates are in fact monomials,
\begin{align}
&p_{123} = 1\,, 
&&p_{234} = 1\,, 
&&p_{345} = x_{11} x_{21}\,, \notag \\
&p_{456} = x_{11}^2 x_{21} x_{12} x_{22}\,,
&&p_{156} = x_{11} x_{12} \,, 
&&p_{126}  = 1\,.
\end{align}
The remaining $\mathcal{A}$-coordinates of the initial cluster and their cyclic images are
\begin{align}
&p_{124}  = 1\,, 
&&p_{235} = 1+x_{11} +x_{11} x_{21}\,, 
&&p_{346} = x_{11} x_{21}(1+x_{12} +x_{12} x_{22})\,, \notag \\
&p_{145} = x_{11} \,, 
&&p_{256} = x_{11} x_{12}(1+x_{21}+x_{21}x_{22})\,, 
&&p_{136} = 1+ x_{11} + x_{11} x_{12}\,,
\end{align}
and
\begin{align}
&p_{134} = 1\,, \quad p_{245} = x_{11}(1+x_{21})\,, \quad p_{356} = x_{11} x_{21}x_{12} (1+x_{22} +x_{11} x_{22})\,, \notag \\
&p_{146} = x_{11}(1+x_{12}) \,, \! \quad  \qquad \qquad \qquad p_{125} = 1\,, \notag \\
&p_{236} = 1+ x_{11} + x_{11} x_{12} + x_{11} x_{21} + x_{11} x_{21} x_{12} + x_{11} x_{21} x_{12} x_{22}\,.
\end{align}
There are two remaining minors which appear as the central nodes of $D_4$-shaped clusters
\begin{align}
p_{135}  &= 1+x_{11}\,, \notag \\
p_{246}  &= x_{11}(1+x_{12}+x_{21}+x_{12} x_{21} + x_{12} x_{21} x_{22})\,.
\label{minors135246}
\end{align}
In addition we have the two quadratic $\mathcal{A}$-coordinates (which also appear in the central node of $D_4$-shaped clusters),
\begin{align}
q_1 = \langle 12[34]56 \rangle &= x_{11} x_{12} x_{21}(1 + x_{22})\,, \notag \\
q_2 = \langle 23[45]61 \rangle &= x_{11}(1+x_{11} + x_{11} x_{12} + x_{11} x_{21} + x_{11} x_{12} x_{21})\,,
\label{quadraticAs}
\end{align}
where $\langle ab[cd]ef\rangle = \langle abd \rangle \langle cef \rangle - \langle abc \rangle \langle def \rangle$ obeys the symmetry properties $\langle ab[cd]ef \rangle = \langle cd[ef]ab \rangle = - \langle cd[ab]ef \rangle$. Under a cyclic transformation $q_1 \rightarrow q_2$ and $q_2 \rightarrow q_1$.

With the expressions of the $\mathcal{A}$-coordinates to hand we may now define a number of different tropical fans. The fan defined by Speyer and Williams to describe the positive part of the tropical Grassmannian is obtained by replacing the polynomial expressions for the Pl\"ucker coordinates $p_{ijk} = \langle ijk \rangle$ by their tropical counterparts $\tilde{p}_{ijk}$. In other words we replace addition with minimum and multiplication with addition. For example, the tropical versions of the minors (\ref{minors135246}) are
\begin{align}
\tilde{p}_{135} &= {\rm min}(0,\tilde{x}_{11})\, ,\notag \\
\tilde{p}_{246}&= {\rm min}(\tilde{x}_{11},\tilde{x}_{11}+\tilde{x}_{12},\tilde{x}_{11}+\tilde{x}_{21},\tilde{x}_{11}+\tilde{x}_{12}+\tilde{x}_{21},\tilde{x}_{11}+\tilde{x}_{12}+\tilde{x}_{21}+\tilde{x}_{22})\, ,
\end{align}
where we have used the notation $\tilde{x}$ to remind the reader that these are tropical counterparts to the original polynomials.
Each tropically evaluated minor defines distinct regions of piecewise linearity. For example, the regions of piecewise linearity of the tropical minor
\be
\tilde{p}_{346} = {\rm min}(\tilde{x}_{11}+\tilde{x}_{21},\tilde{x}_{11}+\tilde{x}_{21}+\tilde{x}_{12},\tilde{x}_{11}+\tilde{x}_{21}+\tilde{x}_{12}+\tilde{x}_{22})\,,
\ee 
are separated by the following tropical hypersurfaces,
\begin{align}
\tilde{x}_{12} &= 0 \leq  \tilde{x}_{22}\,,\notag \\
\text{or } \tilde{x}_{12} + \tilde{x}_{22} &= 0 \leq \tilde{x}_{12} \,, \notag \\
\text{or } \tilde{x}_{12} + \tilde{x}_{22} &= \tilde{x}_{12}  \leq 0\,.
\end{align}

Note that the tropically evaluated frozen variables are simply linear (instead of piecewise linear) as the frozen minors are expressed as monomials in terms of the $\mathcal{X}$-coordinates. Taking all minors together defines the fan of Speyer and Williams \cite{2003math.....12297S}. More precisely, each tropical minor defines a fan via the boundaries of its distinct regions of piecewise linearity. The Speyer-Williams fan is then the common refinement of all the fans defined by the set of tropical minors. The maximal cones of the fan are four-dimensional regions in the $\tilde{x}$ space in which all minors are linear. The intersection of each maximal cone with the unit sphere is a three-dimensional facet of some polyhedral complex. As described in \cite{2003math.....12297S} there are 48 facets of which 46 are tetrahedra and 2 are bipyramids. The facets have boundaries where some minor is between two different regions of linearity. Such boundaries are of dimension two and in this case there are 98 of them and they are all triangles. The triangles themselves are bounded by edges of dimension one, with 66 edges in total. The edges are then bounded by points (corresponding to intersections of rays of the fan with the unit sphere). The ${\rm Gr}(3,6)$ Speyer-Williams fan has the following 16 rays (with the coordinates ordered as $(\tilde{x}_{11},\tilde{x}_{21},\tilde{x}_{12},\tilde{x}_{22})$),
\begin{align}
&(1,0,0,0), &&(-1,0,0,0), &&(1,-1,0,0), &&(0,0,1,-1), \notag \\
&(0,1,0,0), &&(0,-1,0,0), &&(1,0,-1,0), &&(-1,0,0,1), \notag \\
&(0,0,1,0), &&(0,0,-1,0), &&(1,0,0,-1), &&(0,1,1,-1),\notag \\
&(0,0,0,1), &&(0,0,0,-1), &&(0,1,0,-1), &&(1,-1,-1,0)  \,.
\end{align}

The most important point we wish to make here is that we can consider other tropical fans closely related to the Speyer-Williams fan. Firstly we may refine the fan by tropically evaluating also the quadratic $\mathcal{A}$-coordinates (\ref{quadraticAs}),
\begin{align}
\tilde{q}_1 &= {\rm min}(\tilde{x}_{11}+\tilde{x}_{12} +\tilde{x}_{21},\tilde{x}_{11}+\tilde{x}_{12}+\tilde{x}_{21}+\tilde{x}_{22})\,, \notag \\
\tilde{q}_2 &= {\rm min}(\tilde{x}_{11},2\tilde{x}_{11},2\tilde{x}_{11}+\tilde{x}_{12},2\tilde{x}_{11}+\tilde{x}_{21},2\tilde{x}_{11}+\tilde{x}_{12}+\tilde{x}_{21})\,.
\end{align}
Alternatively we can make a less refined fan by not considering the minors $\langle 135 \rangle$ and $\langle 246 \rangle$. One motivation for considering these different fans is that the fan of Speyer and Williams breaks a discrete symmetry of the ${\rm Gr}(3,6)$ (or $D_4$) cluster algebra while both the more refined one and the less refined one manifest it.

More generally we use the notation $F(\mathcal{S})$ to denote the fan obtained by considering the tropical evaluation of a set of $\mathcal{A}$-coordinates $\mathcal{S}$. As we have seen only unfrozen $\mathcal{A}$-coordinates are relevant in defining the fan since the frozen coordinates are all monomials in the $\mathcal{X}$-coordinates and therefore they do not produce tropical hypersurfaces. The three fans we consider in the context of ${\rm Gr}(3,6)$, from least refined to most refined are recorded in Table \ref{Gr36fans} along with their $f$-vectors. As we will describe in more detail below, the most refined fan (the third in the Table \ref{Gr36fans}) is the dual of the $D_4$ cluster polytope and it is simplicial. For this reason we sometimes refer to it as the `cluster fan'. The Speyer-Williams fan is not simplicial in that two pairs of tetrahedra from the cluster fan have been combined into bipyramids. The least refined fan then has two more pairs of tetrahedra combined into bipyramids. When two tetrahedra are combined into a bipyramid, the triangle at the interface is removed. We should stress that the difference between the Speyer-Williams fan and the cluster fan disappears in the case of ${\rm Gr}(2,n)$ since there the minors $\langle ij \rangle$ are the only $\mathcal{A}$-coordinates.

\begin{table}[h]
\centering
\begin{tabular}{c|ccc}
     \toprule
     $\mathcal{S}$ & $f$-vector & tetrahedra & bipyramids\\
     \midrule
     $\{\langle i\, i+1\, j \rangle\}$ & (16, 66, 96, 46) & 42 & 4 \\
     $\{\langle ijk \rangle\}$ &(16, 66, 98, 48) & 46 & 2 \\
     $\{\langle ijk \rangle \} \cup \{q_1,q_2\}$ & (16, 66, 100, 50)& 50& 0 \\
     \bottomrule
\end{tabular}
\caption{Different possible fans for ${\rm Gr}(3,6)$ with their $f$-vectors as well as a characterisation of the dimension two faces.}
\label{Gr36fans}
\end{table}

Another way of encoding all the data listed in Table \ref{Gr36fans} is to split up the $f$-vectors for the three fans as follows,
\begin{equation*}
\begin{aligned}
\{& 16_{1}, 66_2, 96_3, 42_4+4_5 \},\\
\{& 16_1, 66_2,98_3, 46_4+2_5 \} ,\\
\{& 16_1, 66_2, 100_3, 50_4\}.
\end{aligned}
\end{equation*}
Here the subscript notation refers to the number of vertices of each component, i.e. the $2_5$ in the final entry of the middle vector refers to the two five-vertex bipyramids, while the $46_4$ refers to the 46 tetrahedra.

All three fans described above share the same set of 16 rays (and also the same set of 66 edges between rays). As we described in \cite{Drummond:2019qjk} the rays may be obtained as ${\bf g}$-vectors from the associated cluster algebra. Each ${\bf g}$-vector is associated to a cluster $\mathcal{A}$-coordinate, which in turn is associated to a codimension one subalgebra and hence a codimension one boundary of the cluster polytope (see e.g. discussions in \cite{Golden:2013xva,Drummond:2017ssj,Drummond:2018dfd}). This implies that the above fans should all be interpreted as duals of polytopes related to the cluster polytope. In the final case (the cluster fan) the corresponding polytope is exactly the $D_4$ cluster polytope whose codimension one (i.e dimension three) boundary components correspond to the vertices arising from the rays of the fan. The codimension one boundary components are either 14-vertex Stasheff polytopes or 8-vertex cubes. The edges of the fan correspond to intersections of the polytope boundary components and are either 5-vertex pentagons or 4-vertex squares. The triangles of the fan correspond to edges in the polytope and the facets  correspond to vertices of the polytope which correspond to individual clusters of the $D_4$ cluster algebra. The fact that the facets of the cluster fan are all tetrahedra corresponds to the fact that the clusters of the $D_4$ cluster algebra all have four active nodes. 

\begin{figure}
	\centering
	{\footnotesize
  \begin{subfigure}{0.4\textwidth}
    \begin{tikzpicture}[scale=0.9]
    \draw[thick] (-2,2) node[above left] {{\tiny $(0,1,1,-1)$}} -- ($(0,2)+(0.1,0.9)$);
    \draw[thick] (-2,2) -- ($(0,2)+(0.1,-0.9)$);
    \draw[dashed] (-2,2) -- ($(0,2)+(-0.3,0)$);

    \draw[line width=0, fill=gray, opacity=0.2] ($(0,2)+(0.1,0.9)$) -- ($(0,2)+(0.1,-0.9)$) -- ($(0,2)+(-0.3,0)$) -- cycle;
    \draw[thick] ($(0,2)+(0.1,0.9)$) -- ($(0,2)+(0.1,-0.9)$);
    \draw[dashed] ($(0,2)+(0.1,0.9)$) -- ($(0,2)+(-0.3,0)$);
    \draw[dashed] ($(0,2)+(0.1,-0.9)$) -- ($(0,2)+(-0.3,0)$);

    \draw[thick] (2,2) -- ($(0,2)+(0.1,0.9)$);
    \draw[thick] (2,2) -- ($(0,2)+(0.1,-0.9)$);
    \draw[dashed] (2,2) -- ($(0,2)+(-0.3,0)$);
    
    \draw[fill=black] (2,2) circle (0.07);

    
    \draw[thick] (-2,-2) -- ($(0,-2)+(0.1,0.9)$);
    \draw[thick] (-2,-2) -- ($(0,-2)+(0.1,-0.9)$);
    \draw[dashed] (-2,-2) -- ($(0,-2)+(-0.3,0)$);

    \draw[line width=0, fill=gray, opacity=0.2] ($(2,0)+(0.9,0.1)$) -- ($(2,0)+(-0.9,0.1)$) -- ($(2,0)+(0,-0.3)$) -- cycle;
    \draw[thick] ($(0,-2)+(0.1,0.9)$) -- ($(0,-2)+(0.1,-0.9)$);
    \draw[dashed] ($(0,-2)+(0.1,0.9)$) -- ($(0,-2)+(-0.3,0)$);
    \draw[dashed] ($(0,-2)+(0.1,-0.9)$) -- ($(0,-2)+(-0.3,0)$);

    \draw[thick] (2,-2) node[below right] {{\tiny $(1,-1,-1,0)$}} -- ($(0,-2)+(0.1,0.9)$);
    \draw[thick] (2,-2) -- ($(0,-2)+(0.1,-0.9)$);
    \draw[dashed] (2,-2) -- ($(0,-2)+(-0.3,0)$);

     \draw[fill=black] (2,-2) circle (0.07);

    
    \draw[thick] (2,2) node[above right] {{\tiny $(1,0,0,-1)$}} -- ($(2,0)+(0.9,0.1)$);
    \draw[thick] (2,2) -- ($(2,0)+(-0.9,0.1)$);
    \draw[thick] (2,2) -- ($(2,0)+(0,-0.3)$);

    \draw[line width=0, fill=gray, opacity=0.2] ($(0,-2)+(0.1,0.9)$) -- ($(0,-2)+(0.1,-0.9)$) -- ($(0,-2)+(-0.3,0)$) -- cycle;
    \draw[dashed] ($(2,0)+(0.9,0.1)$) -- ($(2,0)+(-0.9,0.1)$);
    \draw[thick] ($(2,0)+(-0.9,0.1)$) -- ($(2,0)+(0,-0.3)$);
    \draw[thick] ($(2,0)+(0,-0.3)$) --($(2,0)+(0.9,0.1)$);
    
    \draw[thick] (2,-2) -- ($(2,0)+(0.9,0.1)$);
    \draw[thick] (2,-2) -- ($(2,0)+(-0.9,0.1)$);
    \draw[thick] (2,-2) -- ($(2,0)+(0,-0.3)$);
    
     \draw[fill=black] (-2,2) circle (0.07);

    
    \draw[thick] (-2,2) -- ($(-2,0)+(0.9,0.1)$);
    \draw[thick] (-2,2) -- ($(-2,0)+(-0.9,0.1)$);
    \draw[thick] (-2,2) -- ($(-2,0)+(0,-0.3)$);

    \draw[line width=0, fill=gray, opacity=0.2] ($(-2,0)+(0.9,0.1)$) -- ($(-2,0)+(-0.9,0.1)$) -- ($(-2,0)+(0,-0.3)$) -- cycle;
    \draw[dashed] ($(-2,0)+(0.9,0.1)$) -- ($(-2,0)+(-0.9,0.1)$);
    \draw[thick] ($(-2,0)+(-0.9,0.1)$) -- ($(-2,0)+(0,-0.3)$);
    \draw[thick] ($(-2,0)+(0,-0.3)$) --($(-2,0)+(0.9,0.1)$);
    
    \draw[thick] (-2,-2) node[below left] {{\tiny $(-1,0,0,1)$}} -- ($(-2,0)+(0.9,0.1)$);
    \draw[thick] (-2,-2) -- ($(-2,0)+(-0.9,0.1)$);
    \draw[thick] (-2,-2) -- ($(-2,0)+(0,-0.3)$);

    \draw[fill=black] (-2,-2) circle (0.07);
   
  \end{tikzpicture}
		\label{fig:tr25}
		\end{subfigure}
              }
              \quad\quad\quad
		{\footnotesize
		  \begin{subfigure}{0.4\textwidth}
                    \begin{tikzpicture}[scale=1]
\pgfmathsetmacro{\aa}{3}
\pgfmathsetmacro{\bb}{1.8}

\draw[thick,color=gray] (0,0) -- (0,-\aa+\bb);
\draw[thick,color=gray] (1.8,1.8) -- (1.8+\aa-\bb,1.8);
\draw[thick,color=gray] (\aa+\bb,0) -- (\aa+\bb,-\aa+\bb);
\draw[thick,color=gray] (1.8,-\aa+\bb-1.8) -- (1.8+\aa-\bb,-\aa+\bb-1.8);

\draw[thick] (0,0) -- (1.1,0.1) -- (1.2,1.2) -- (0.1,1.1) node[above left] {{\scriptsize $q_1$}}  -- cycle;
\draw[thick] (1.1,0.1) -- (1.7,0.7) -- (1.8,1.8) -- (0.7,1.7) -- (0.1,1.1);
\draw[thick] (1.2,1.2) -- (1.8,1.8);
\draw[thick] (0,0) -- (0.55,0.55);
\draw[thick] (0.7,1.7) -- (0.55,0.55) -- (1.7,0.7);
\draw[fill=black] (0,0) circle (0.07);
\draw[fill=black] (1.8,1.8) circle (0.07);

\draw[thick] (-0+\aa+\bb,0) -- (-1.1+\aa+\bb,0.1) -- (-1.2+\aa+\bb,1.2) -- (-0.1+\aa+\bb,1.1)  -- cycle;
\draw[thick] (-1.1+\aa+\bb,0.1) -- (-1.7+\aa+\bb,0.7) -- (-1.8+\aa+\bb,1.8) -- (-0.7+\aa+\bb,1.7) -- (-0.1+\aa+\bb,1.1) node[above right] {{\scriptsize $\langle 246 \rangle$}};
\draw[thick] (-1.2+\aa+\bb,1.2) -- (-1.8+\aa+\bb,1.8);
\draw[thick] (-0+\aa+\bb,0) -- (-0.55+\aa+\bb,0.55);
\draw[thick] (-0.7+\aa+\bb,1.7) -- (-0.55+\aa+\bb,0.55) -- (-1.7+\aa+\bb,0.7);
\draw[fill=black] (-1.8+\aa+\bb,1.8) circle (0.07);
\draw[fill=black] (-0+\aa+\bb,0) circle (0.07);

\draw[thick] (0+\aa,0-\aa) -- (1.1+\aa,0.1-\aa) -- (1.2+\aa,1.2-\aa) -- (0.1+\aa,1.1-\aa)  -- cycle;
\draw[thick] (1.1+\aa,0.1-\aa) -- (1.7+\aa,0.7-\aa) node[below right] {{\scriptsize $q_2$}} -- (1.8+\aa,1.8-\aa) -- (0.7+\aa,1.7-\aa) -- (0.1+\aa,1.1-\aa);
\draw[thick] (1.2+\aa,1.2-\aa) -- (1.8+\aa,1.8-\aa);
\draw[thick] (0+\aa,0-\aa) -- (0.55+\aa,0.55-\aa);
\draw[thick] (0.7+\aa,1.7-\aa) -- (0.55+\aa,0.55-\aa) -- (1.7+\aa,0.7-\aa);
\draw[fill=black] (0+\aa,0-\aa) circle (0.07);
\draw[fill=black] (1.8+\aa,1.8-\aa) circle (0.07);

\draw[thick] (0+\bb,0-\aa) -- (-1.1+\bb,0.1-\aa)  -- (-1.2+\bb,1.2-\aa) -- (-0.1+\bb,1.1-\aa) -- cycle;
\draw[thick] (-1.1+\bb,0.1-\aa) -- (-1.7+\bb,0.7-\aa) node[below left] {{\scriptsize $\langle 135 \rangle$}} -- (-1.8+\bb,1.8-\aa) -- (-0.7+\bb,1.7-\aa) -- (-0.1+\bb,1.1-\aa);
\draw[thick] (-1.2+\bb,1.2-\aa) -- (-1.8+\bb,1.8-\aa);
\draw[thick] (-0+\bb,0-\aa) -- (-0.55+\bb,0.55-\aa);
\draw[thick] (-0.7+\bb,1.7-\aa) -- (-0.55+\bb,0.55-\aa) -- (-1.7+\bb,0.7-\aa);
\draw[fill=black] (0,-\aa+\bb) circle (0.07);
\draw[fill=black] (1.8,-\aa+\bb-1.8) circle (0.07);

\end{tikzpicture}
 \label{}
	\end{subfigure}
	}
	\caption{\small Left: a subset of the cluster fan (or more precisely its intersection with the unit sphere) showing eight tetrahedral facets and four highlighted triangles. The highlighted vertices correspond to the four rays given. Right: the subgraph of the $D_4$ cluster polytope formed by keeping only the clusters whose active nodes are connected in the shape of a $D_4$ Dynkin diagram. The cubes correspond to the $\mathcal{A}$-coordinates shown and are dual to the four highlighted vertices of the left figure. The grey edges connecting the four cubes are dual to the highlighted triangles of the left figure. The highlighted vertices (clusters) are dual to the eight tetrahedra of the left figure.}
	\label{D4clusterfanpoly}
\end{figure}
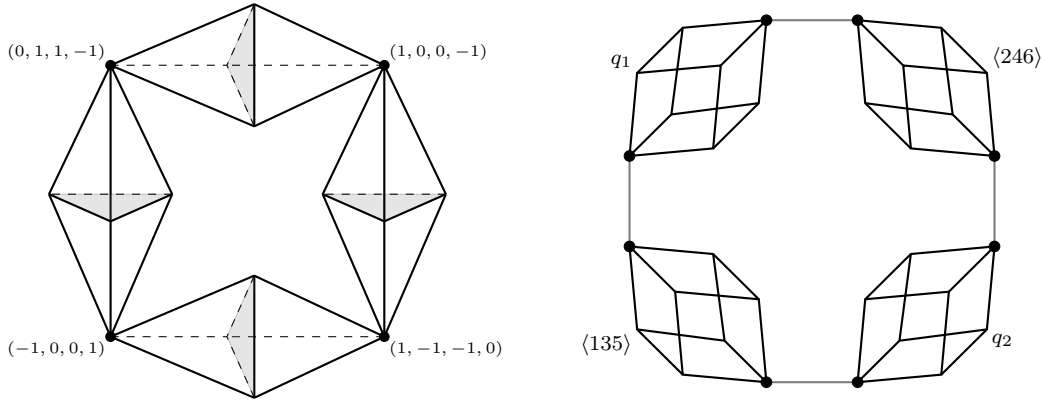

In Fig. \ref{D4clusterfanpoly} we illustrate relevant parts of the cluster fan and its dual $D_4$ cluster polytope. The left figure shows all sixteen rays but only eight of the tetrahedal facets. The facets shown come in pairs in which the two tetrahedra intersect on a common triangle. The right figure shows the connectivity of the subset of clusters in the $D_4$ cluster polytope which have the topology of the $D_4$ Dynkin diagram. Only the four codimension one boundaries with the topology of cubes are therefore shown fully. The cubes are dual to the rays at the four marked corners of the left figure. Each cube is connected to its two neighbours by a single edge, dual to the corresponding shaded triangle in the left figure.

The polytope dual to the Speyer-Williams fan can be obtained from the $D_4$ cluster polytope by shrinking the two vertical grey edges connecting the cubes in the right half of Fig \ref{D4clusterfanpoly} so that two pairs of vertices become six-valent. These then correspond to the two bipyramidal facets of the Speyer-Williams fan. To obtain the polytope dual to the first fan listed in Table \ref{Gr36fans} one should also shrink the two horizontal grey edges connecting corners of distinct cubes. The edges which are shrunk in this procedure therefore do not belong to the cubes but only the Stasheff polytopes and moreover only belong to pentagonal faces, not square faces. When an edge is shrunk the corresponding dual triangle is deleted from the fan and the associated pair of tetrahedra combine into a bipyramid. The triangles that can be removed in this way have a special property: the vertices are all \emph{disconnected} neighbours in the sense of \cite{Drummond:2017ssj}. That is the $\mathcal{A}$-coordinates associated to any pair do appear in clusters together, but never connected by an arrow of the quiver diagram.

Finally let us point out that there is also a fan, topologically equivalent to the Speyer-Williams fan, which can be obtained by taking $\mathcal{S}$ to be the set of all $\mathcal{A}$-coordinates, except $\langle 135 \rangle$ and $\langle 246 \rangle$. The dual polytope to this fan would the simply be obtained from the cluster polytope by shrinking only the horizontal grey edges of Fig. \ref{D4clusterfanpoly} and not the vertical ones.

The above statements can also be encoded in the splitting of the $f$-vectors of the dual polytopes as follows,
\begin{equation*}
\begin{aligned}
\{& 46_{1}, 96_2, 42_4 + 24_5, 4_8 + 12_{13} \},\\
\{& 48_1, 98_2, 36_4+30_5,4_8+6_{13}+6_{14} \},\\
\{& 50_1, 100_2, 30_4+36_5, 4_8+12_{14} \}.
\end{aligned}
\end{equation*}
Here the final line again corresponds to the cluster polytope with the $12_{14}$ in the final line referring to the twelve Stasheff polytope codimension-one boundaries and the $4_8$ the four cubes. After shrinking the first pair of edges we obtain the middle line (dual to the Speyer-Williams fan) where the 50 vertices have become 48, the two shrunk edges are missing leaving only 98 of the original 100, 6 of the 36 pentagons have become squares and 6 of the Stasheff polytopes have been shrunk to 13-vertex objects.

It is clear that the relation between Grassmannian cluster algebras and tropical Grassmannians is really related to a whole family of possible fans and their dual polytopes with the cluster fan being the most refined and the dual of the cluster polytope. The other fans, including that of Speyer and Williams describing the positive part of the tropical Grassmannian are obtained by shrinking edges in the cluster polytope or equivalently removing data from the corresponding tropical fan. As we will see in the next section, this will lead to a generalised set of scattering equations associated to each fan.

We may similarly describe different fans in the case of
${\rm Gr}(3,7)$. The fans and their dual polytopes are harder to
picture but we can describe the relevant features by means of the
notation introduced above. Let us first introduce a notation for the
42 unfrozen $\mathcal{A}$-coordinates of the ${\rm Gr}(3,7)$ cluster
algebra which come in six cyclic classes,
\begin{align}
&a_{11} = \langle 347 \rangle,
&&a_{21} = \langle 134 \rangle, 
&&a_{31} = \langle 156 \rangle, \notag \\
&a_{41} = \langle 257 \rangle,
&&a_{51} = \langle 12[34]56 \rangle,
&&a_{61} = \langle 61[23]45 \rangle\,.
\end{align}
The remaining $a_{ij}$ are obtained by cyclic rotations of the above.\footnote{The notation has been chosen to match existing notation on $\mathcal{A}$-coordinates for ${\rm Gr}(4,7)$ which we will study further in Sect. \ref{polylogs}.}

The fans we wish to consider in this case are given by choosing four possibilities for the set $\mathcal{S}$ of $\mathcal{A}$-coordinates as described in Table \ref{Gr37fans}. The second fan in Table \ref{Gr37fans} corresponds to the Speyer-Williams fan since the set $\{a_{1i},a_{2i},a_{3i},a_{4i}\}$ corresponds to taking all minors to define the fan. There is one less refined fan where we omit the $a_{4i}$ from $\mathcal{S}$. There are two more refined fans, one where we include also the $a_{5i}$ and a final one where we take all $\mathcal{A}$-coordinates, which we will again refer to as the cluster fan. 

Once again the cluster fan ($\{a_{1i} ,\ldots , a_{6i}\}$) is simplicial. To obtain the $\{a_{1i}, \ldots, a_{5i}\}$ fan, seven triangles are removed, indicating the presence of seven bipyramids in dimension three. To then obtain the Speyer-Williams fan ($\{a_{1i}, \ldots ,a_{4i}\}$) seven edges are removed indicating that seven pairs of triangles combine into squares. The remaining triangles (49 of them) including these missing edges are then removed as well as 21 further triangles which do not involve the removed edges, leaving 1456 triangles and 7 squares in dimension two. To obtain the least refined fan ($\{a_{1i}, \ldots , a_{3i}\}$) one then removes a further 7 edges meaning another seven pairs of triangles combine into squares and another 42 triangles involving the removed edges are lost. In addition a further 21 triangles are removed leaving 1379 triangles and 14 squares in dimension two.
\begin{table}[h]
\centering
\begin{tabular}{c|ccc}
     \toprule
     $\mathcal{S}$ & $f$-vector & triangles & squares\\
     \midrule
     $\{a_{1i},a_{2i},a_{3i}\}$ & (42, 385, 1393, 2373, 1918, 595) & 1379 & 14 \\
     $\{a_{1i},a_{2i},a_{3i},a_{4i}\}$ & (42, 392, 1463, 2583, 2163, 693) & 1456 & 7 \\
     $\{a_{1i},a_{2i},a_{3i},a_{4i},a_{5i}\}$ & (42, 399, 1540, 2821, 2443, 805) & 1540 & 0 \\
     $\{a_{1i},a_{2i},a_{3i},a_{4i},a_{5i},a_{6i}\}$ & (42, 399, 1547, 2856, 2499, 833) & 1547 & 0 \\
     \bottomrule
\end{tabular}
\caption{Different possible fans for ${\rm Gr}(3,7)$ with their $f$-vectors as well as a characterisation of the dimension two faces.}
\label{Gr37fans}
\end{table}

Using the same notation introduced in ${\rm Gr}(3,6)$ we can compactly include the above information and more into refined $f$-vectors which split up repsectively as,
\begin{equation}
\begin{aligned}
\{&42_1, 385_2, 1379_3+14_4,2240_4+133_5, 1659_5+196_6+63_7, 455_6+84_7+28_8+28_9\},\\
\{&42_1, 392_2, 1456_3+7_4, 2506_4+77_5,1995_5+140_6+28_7, 595_6+63_7+28_8+7_9\},\\
\{&42_1, 399_2, 1540_3, 2814_4+7_5, 2415_5+28_6,777_6+28_7\},\\
\{&42_1,399_2,1547_3,2856_4,2499_5,833_6\}.
\end{aligned}
\end{equation}
Here we remind the reader that the notation $n_m$ means $n$ faces, each one consisting of $m$ vertices (rays), with the dimension of the face increasing from zero to five as we proceed from left to right along the $f$-vector.

Also as in the ${\rm Gr}(3,6)$ case we can think of all of the fans as being dual to polytopes. The cluster fan is dual to the ${\rm Gr}(3,7)$ (or $E_6$) cluster polytope. The other polytopes are then successively obtained by shrinking edges in this polytope. We can capture a lot of information about the shrinking by splitting the $f$-vectors of the dual polytopes which are respectively,
\begin{equation*}
\begin{aligned}
\{&595_1, 1918_2, 1848_4 + 525_5, 651_8 + 448_{10} + 14_{12} + 252_{13} + 28_{14},\\
&91_{16}+70_{20}+7_{25}+98_{26}+14_{33}+42_{34}+28_{37}+21_{38}+14_{46},\\
&7_{32}+14_{68}+7_{98}+14_{138} \},\\
\{&693_1, 2163_2, 1841_4+742_5, 525_8+574_{10}+7_{12}+161_{13}+196_{14}, \\
& 42_{16}+ 112_{20}+ 7_{25}+ 49_{26}+ 49_{28}+ 7_{33}+ 7_{34}+ 56_{37}+ 14_{40}+ 28_{42}+ 14_{48}+ 7_{49}, \\
& 7_{40}+ 14_{74}+ 7_{112}+ 7_{144}+ 7_{170}\},\\
\{&805_1, 2443_2, 1834_4 + 987_5, 406_8 + 658_{10} + 28_{12} + 84_{13} + 364_{14},\\
&7_{16}+112_{20}+14_{24}+21_{25}+28_{26}+70_{28}+56_{40}+56_{42}+7_{48}+14_{49}+14_{50},\\
&7_{50}+14_{80}+7_{128}+14_{178}\},\\
\{&833_1, 2499_2, 1785_4+1071_5, 357_8+714_{10}+476_{14}, \\ &119_{20}+21_{25}+112_{28}+112_{42}+35_{50}, 7_{50}+14_{84}+7_{132}+14_{182}\}.
\end{aligned}
\end{equation*}
As an example of the information captured in the above splittings, we see in the final entry of the final line the codimension one subalgebras of the $E_6$ polytope: with $7_{50}$ corresponding to the 7 $A_2 \times A_2 \times A_1$ subalgebras, $14_{84}$ to the 14 $A_4 \times A_1$, $7_{132}$ to the 7 $A_5$ and $14_{182}$ to the 14 $D_5$.

Finally, we can give some information on the structure of various fans in the infinite case ${\rm Gr}(4,8)$ studied in recent papers \cite{Drummond:2019qjk,Drummond:2019cxm,Arkani-Hamed:2019rds,Henke:2019hve}. In the infinite case there is no cluster fan as there are infinitely many $\mathcal{A}$-coordinates. Moreover, as discussed in \cite{Drummond:2019cxm,Arkani-Hamed:2019rds,Henke:2019hve} there are additional rays which are not obtained from any ${\bf g}$-vector of the cluster algebra but rather arise as limits of infinite sequences of rays. Nevertheless we can systematically solve the tropical conditions for the 3 different fans in ${\rm Gr}(4,8)$ considered in \cite{Drummond:2019cxm}. These fans were the Speyer-Williams fan, generated by taking $\mathcal{S}$ as the set of all minors, a reduced fan, obtained by taking the maximal parity-invariant subset of minors, and an augmented fan, obtained by taking the parity completion of all minors. We find for their $f$-vectors,
\begin{equation*}
\begin{aligned}
f_{48, \rm{red}} &= (274,5782,46312,189564,447284,635176,536960,249306,49000),\\
f_{48, \rm{SW}} &=(360, 7984, 66740, 285948, 706042, 1047200, 922314, 444930, 90608),\\
f_{48, \rm{aug}} &=(548, 12748, 111104, 492548, 1251188, 1900152, 1706592, 836570, 172588).
\end{aligned}
\end{equation*}
The maximal cones of the three fans split as
\begin{equation*}
\begin{aligned}
49000&=22636_{9}+7872_{10}+4728_{11}+4528_{12}+2048_{13}+2544_{14}+960_{15}+672_{16} \\
&~ +1488_{17}+664_{18}+232_{19}+128_{20}+128_{21}+128_{22}+32_{23}+64_{24}+48_{25} \\
&~ +64_{28}+32_{34}+4_{45},\\
90608&=50356_{9}+12320_{10}+9116_{11}+6064_{12}+4448_{13}+2332_{14}+2176_{15}+872_{16}\\
&~ +976_{17}+676_{18}+384_{19}+336_{20}+200_{21}+48_{22}+8_{23}+80_{24}+72_{25}+24_{26}\\
&~ +48_{27}+16_{29}+20_{33}+16_{34}+16_{36}+4_{49},\\
172588&=112708_{9}+21008_{10}+13088_{11}+10016_{12}+4480_{13}+3440_{14}+2272_{15}\\
&~+1184_{16}+1888_{17}+1168_{18}+336_{19}+160_{20}+256_{21}+192_{22}+48_{23}+128_{24}\\
&~+80_{25}+64_{28}+32_{32}+32_{34}+8_{45}.
\end{aligned}
\end{equation*}

In the above computations, the computer package {\tt polymake} \cite{polymake:2000} was used.

\section{Generalised scattering equations}

In \cite{Cachazo:2019ngv} Cachazo, Early, Guevara and Mizera proposed a relation between the tropical Grassmannians ${\rm Gr}(k,n)$ and a set of scattering equations which generalise the scattering equations introduced in \cite{Cachazo:2013gna,Cachazo:2013hca,Cachazo:2013iea} for ${\rm Gr}(2,n)$. Here we would like to emphasise the point that there is a set of generalised scattering equations for each choice of tropical fan $F(\mathcal{S})$ described in the previous section, with the equations of \cite{Cachazo:2019ngv} corresponding to the Speyer-Williams fans. In the finite cases, the most refined fan (the cluster fan) is associated to the most general set of scattering equations.

Let us first review the scattering equations of \cite{Cachazo:2019ngv} before introducing their generalisations. 
One starts with the potential function
\begin{equation}\label{potential}
F=\sum_{i_1 <i_2 <\ldots < i_k } s_{i_1 i_2 \ldots i_k} \log \langle i_1 i_2   \ldots i_k \rangle\,.
\end{equation}
Here $\langle i_1 i_2   \ldots i_k \rangle $ are minors of the Grassmannian $(k,n)$ matrix which depend on $(k-1)(n-k-1)$ variables\footnote{For example by choosing coordinates via the web matrix $W$ defined in (\ref{webmatrix}).}, and $s_{i_1 i_2 \ldots i_k}$  are generalised Mandelstam variables, totally symmetric in their $k$ indices. The generalised Mandelstam variables satisfy generalised momentum conservation relations
\begin{equation}\label{momentum_conservation}
\sum_{i_2  <\ldots < i_k} s_{i_1 i_2  \ldots i_k} = 0, \quad \forall~ i_1.
\end{equation}
The generalised momentum conservation relations guarantee the homogeneity of the potential $F$ under the rescalings of the $n$ columns of the $k \times n$ matrix.
The scattering equations are then defined to be 
\begin{equation}\label{scattering equations}
d F = 0\,.
\end{equation}
These equations are to be interpreted as equations for the coordinates parametrising the matrix (e.g. the cluster $\mathcal{X}$ coordinates) in terms of the generalised Mandelstam variables $s_{i_1 \ldots i_k}$.
The $\phi^3$ amplitude is then evaluated as a localised integral of Parke-Taylor factors (see \cite{Cachazo:2019ngv}). 

The $\phi^3$ amplitude thus obtained can also be identified with the volume of the fan (or its intersection with the unit sphere), which itself can be computed by triangulating and adding the volume of all simplicial facets \cite{Drummond:2019qjk}. This picture generalises the kinematic associahedron picture of \cite{Arkani-Hamed:2017mur} which computes the volume of the ${\rm Gr}(2,n)$ fan, in which the volume of each facet is simply a tree-level $\phi^3$ Feynman diagram. As we stressed in \cite{Drummond:2019qjk}, in the cases where the Grassmannian cluster algebra is finite, the cluster algebra provides a useful way to immediately obtain a triangulation of the Speyer-Williams fan and thus obtain the amplitude as a function of the generalised Mandelstam invariants $s_{i_1 i_2 \ldots i_k}$ via its volume. 

Also discussed in \cite{Drummond:2019qjk} was the fact that we can recover the rays in the positive part of the tropical Grassmannian of Speyer and Sturmfels \cite{SpeyerSturmfels} by simply evaluating all the tropical minors on the rays of the Speyer-Williams fan. Such rays can be expressed in terms of the generalised Mandelstam invariants if we form the scalar product of the vector of generalised Mandelstam invariants with the vector of tropical minors. For example, in the case of ${\rm Gr}(3,6)$ we have
\be
(\tilde{x}_{11},\tilde{x}_{21},\tilde{x}_{12},\tilde{x}_{22}) \mapsto \sum s_{i_1 i_2 i_3} \tilde{p}_{i_1 i_2 i_3}(\tilde{x}_{11},\tilde{x}_{21},\tilde{x}_{12},\tilde{x}_{22})\,.
\label{SSray}
\ee
More explicitly, if we take the rays describing the vertices of the bipyramid on the left side of the left figure in Fig. \ref{D4clusterfanpoly} and evaluate the quantity (\ref{SSray}) we find
\begin{align}
&(0,0,1,0)&&\mapsto  &&t_{1234}\,, \notag \\
&(-1,0,0,0)&&\mapsto && t_{1256}\,, \notag \\
&(0,1,0,0)&&\mapsto && t_{3456}\,, \notag \\
&(0,1,1,-1)&&\mapsto && r_{123456} \,, \notag \\
&(-1,0,0,1)&&\mapsto && r_{341256} \,.
\end{align}
Here we use the notation 
\begin{align}
t_{ijkl} &= s_{ijk} + s_{ijl} + s_{ikl} + s_{jkl}\,, \notag \\
r_{ijklmn} &= t_{ijkl} + s_{klm} + s_{kln}\,.
\end{align}
The full set of sixteen rays becomes the set of $s_{123}$ (and its five cyclic cousins),  $t_{1234}$ (and its five cyclic cousins), $r_{123456}$ (and one cyclic cousin), $r_{341256}$ (and one cyclic cousin). The fact that these five rays form a bipyramid is reflected in the fact that the Mandelstam form of the rays obeys
\be
t_{1234} + t_{1256} + t_{3456} = r_{123456} + r_{341256}\,.
\label{bipyramidrel}
\ee
The reverse map from the kinematic expression to the Speyer-Williams ray consists of tropically evaluating the $\mathcal{X}$-coordinates in terms of minors \cite{2003math.....12297S,Drummond:2019qjk}.

The above description corresponds to the choice of the Speyer-Williams fan. To obtain the less refined ${\rm Gr}(3,6)$ fan described in Sect.~\ref{Sect-fans} we simply set $s_{135} = s_{246} = 0$ in the above discussion. This suggests that there should also be a further generalisation of the scattering equations which corresponds the more refined cluster fan. Indeed we propose a further generalisation of the potential function,
\begin{equation}\label{potentialGen}
F=\sum_{i_1 <i_2 <i_3} s_{i_1 i_2 i_3} \log \langle i_1 i_2  i_3 \rangle + s_{q_1} \log q_1 + s_{q_2} \log q_2\,.
\end{equation}
Here we have added two more terms corresponding to the two quadratic $\mathcal{A}$-coordinates (\ref{quadraticAs}) and introduced new generalised Mandelstam variables $s_{q_1}$ and $s_{q_2}$. 
The momentum conservation relation now reads 
\begin{equation}\label{momentum_conservation}
\sum_{i_2 <i_3} s_{i_1 i_2 i_3} + s_{q_1} +  s_{q_2} = 0, \quad \forall~ i_1.
\end{equation}
The relation (\ref{momentum_conservation}) again guarantees the homogeneity of the potential function (\ref{potentialGen}). The scattering equations are the same as in \eqref{scattering equations}. To return to the system corresponding to the Speyer-Williams fan we simply set $s_{q_1} = s_{q_2} = 0$ in the new system. If we do not set $s_{q_1}$ and $s_{q_2}$ to zero then the expressions for the sixteen rays become modified as follows,
\begin{align} 
s_{123} &\mapsto s_{123} \,, \notag \\
t_{1234} &\mapsto t_{1234} + s_{q_1}\,, \notag \\
r_{123456} &\mapsto r_{123456} + s_{q_1} \,, \notag \\
r_{341256} &\mapsto r_{341256} + s_{q_1}\,. 
\end{align}
As before the rays above generated cyclic classes of size six, six, two and two respectively (where $s_{q_1} \rightarrow s_{q_2}$ and $s_{q_2} \rightarrow s_{q_1}$ under a cyclic transformation).

If we perform the above replacements we see that the relation (\ref{bipyramidrel}) will no longer hold since the LHS acquires an additional $3 s_{q_1}$ while the RHS only acquires $2 s_{q_1}$. This is in accordance with the fact that these five rays no longer form a bipyramid in the cluster fan but rather two tetrahedra separated by a triangle.
 
 We can then form a generalised $\phi^3$ amplitude computed from the volume of each facet, just as in the Speyer-Williams case. We obtain a sum over 50 terms (one for each facet - now all tetrahedra) which now depend also on $s_{q_1}$ and $s_{q_2}$. For example, the two tetrahedra described above contribute two of the 50 terms:
 \begin{align}
 &\frac{1}{(t_{1234}+s_{q_1})(t_{1234}+s_{q_1})(t_{1234}+s_{q_1})(r_{123456}+s_{q_1})} \notag \\
 &+ \frac{1}{(t_{1234}+s_{q_1})(t_{1234}+s_{q_1})(t_{1234}+s_{q_1})(r_{341256}+s_{q_1})}\,.
 \end{align}
By construction the amplitude obtained this way reduces to the amplitude of \cite{Cachazo:2019ngv} upon setting $s_{q_1} = s_{q_2} = 0$. The amplitude described above should be equivalent to the lowest order contribution to the integrals over the stringy canonical forms discussed in \cite{Arkani-Hamed:2019mrd}.

In order to justify the equivalence of the generalised scattering equations and tropical fans beyond the Speyer-Williams case considered in \cite{Cachazo:2019ngv}, we consider an example.  We focus on the cluster fan of ${\rm Gr}(3,6)$ and choose the special kinematics where $s_{ijk} = s_{ij}+s_{ik}+s_{jk}$, $s_{i,i+1}=1$, $s_{i,i+2}=-1$.  Analogous kinematics was also considered for the $(2,n)$ case in \cite{Cachazo:2013iea}, where it was shown that each term of the amplitude (each Feynman diagram) contributes 1, thus the scattering amplitude equals the number of all possible diagrams which is the Catalan number.  In ${\rm Gr}(3,6)$ our special kinematics does not have the effect that each of the 50  terms of the amplitude contributes 1, but it does simplify the scattering equations obtained from the potential (\ref{potentialGen}).

Let us choose coordinates for the $(3 \times 6)$ matrix as follows
\begin{equation}
m_{36}=
\left(
\begin{array}{cccccc}
 1 & 0 & 0 & 1 & 1 & 1 \\
 0 & 1 & 0 & 1 & x_5 & x_6 \\
 0 & 0 & 1 & 1 & y_5 & y_6 \\
\end{array}
\right),
\end{equation}
with $\langle ijk \rangle$ now being the minors of $m_{36}$.
The scattering equations to be solved are
\begin{equation}
\frac{\partial F}{\partial x_5}=
\frac{\partial F}{\partial x_6}=
\frac{\partial F}{\partial y_5}=
\frac{\partial F}{\partial y_6}=0\,,
\end{equation}
with $F$ given in (\ref{potentialGen}).
For our chosen kinematics, the generalised momentum  conservation relations imply $s_{q_1}=-s_{q_2} \equiv t$.  The expected amplitude evaluated from adding up the volume of the 50 facets is
\begin{equation}
A_{36} = -
\frac{ 2(3 t^4 -68 t^2 + 288)} {(t^2-4)^3}.
\end{equation} 
Choosing numerical values for $t$ we are able to solve the scattering equations.  Generically, we find 8 solutions.  Then we consider the sum over the solutions of the scattering equations
\begin{equation}
A_{36} = \sum_{\rm slns} 
\frac{1}{\det' \Phi'}
\frac{1}{\prod_{i=1}^6 \langle i\,i+1\,i+2 \rangle},
\end{equation}
where
\begin{equation}
\Phi = \left(
\begin{array}{cc}
 \phi_1 & \phi_3 \\
 \phi_3^{\rm T} & \phi_2
\end{array}
\right),
\qquad 
\phi_1 = \frac{\partial^2 F}{\partial x_a \partial x_b},~
\phi_2 = \frac{\partial^2 F}{\partial y_a \partial y_b},~
\phi_3 = \frac{\partial^2 F}{\partial x_a \partial y_b},
\end{equation}
and $\Phi'$ the matrix $\Phi$ after the removal of rows and columns 1,2,3,4,7,8,9,10.  Explicitly
\begin{equation} 
{\det}' \Phi' = 
\frac{\det \Phi'}
{(\langle 123 \rangle \langle 234 \rangle \langle 341 \rangle \langle 412 \rangle)^2}.
\end{equation}
We have solved the scattering equations for various values of $t$ and found agreement with the expected answer.

The extension of the generalised $\phi^3$ amplitude to more refined fans clearly also generalises to higher ${\rm Gr}(k,n)$. For ${\rm Gr}(3,7)$ one can introduce a new set of Mandelstam variables $s_{q_{5i}} $ corresponding to the $a_{5}$-type quadratic $\mathcal{A}$-coordinates and also $s_{q_{6i}}$ for the $a_{6}$ type. The potential $F$ now reads
\be
F = \sum_{i<j<k} s_{ijk} \log \langle ijk \rangle + \bigl[(s_{q_{51}} \langle 23 [45] 67 \rangle + s_{q_{61}} \langle 56[72]34\rangle) +{\text{cyc.}}\bigr]\,.
\label{gr37genmand}
\ee
The generalised momentum conservation relation reads,
\be
\sum_{j<k} s_{ijk} + \sum_{j\neq i} (s_{q_{5j}} + s_{q_{6j}}) = 0\,.
\label{gr37momcons}
\ee
The above system corresponds to the $\{a_{1i},\ldots,a_{6i}\}$ fan described in Sect. \ref{Sect-fans}. To obtain the $\{a_{1i},\ldots,a_{5i}\}$ fan one simply imposes $s_{q_{6i}}=0$. To obtain the Speyer-Williams fan one imposes also $s_{q_{5i}}=0$. To then obtain the $\{a_{1i},\ldots,a_{3i}\}$ fan one imposes further that $s_{135}=0$ and the cyclically related relations.

One can similarly make a generalisation of the scattering equations corresponding to the ${\rm Gr}(3,8)$ cluster fan (or $E_8$ cluster fan). To do so one needs new Mandelstam variables corresponding to the quadratic and cubic $\mathcal{A}$-coordinates. We will return to this case later. For $k=3$ and $n>8$ there does not exist an analogue of the cluster fan but there are certainly fans which are more refined than the Speyer-Williams fans which therefore introduce new Mandelstam variables beyond the $s_{i_1 \ldots i_k}$.

\section{Cluster polytopes and face variables}

The cluster polytopes can be defined in terms of face variables. Such variables have been discussed in many recent papers \cite{Drummond:2018dfd,Arkani-Hamed:2019mrd,Arkani-Hamed:2019plo} and generalise the dihedral coordinates of ${\rm Gr}(2,n)$ (see e.g. \cite{Brown:2009qja}) to more general cluster polytopes. Face variables have the property that they are valued between $0$ and $1$ in the positive region (which is also the region where all cluster $\mathcal{X}$-coordinates are positive. Each codimension one boundary $a$ of the cluster polytope has an associated face variable $u_a$ and $u_a=0$ defines the boundary. Furthermore on every other codimension one boundary $b$ that does not intersect the defining boundary $a$ the variables $u_a$ takes the value 1.

\begin{figure}
\begin{center}
\begin{subfigure}{0.45\textwidth}
\centering
\begin{tikzpicture}[scale=1.3]
\node[] (x3) at (-1, 0) {$x_3$};
\node[] (x5) at (-2, 0) {$x_5$}; 
\node[] (x2) at (0, 0)  {$x_2$};         
\node[] (x4) at (1, 0)  {$x_4$};        
\node[] (x6) at (2, 0)  {$x_6$};          
\node[] (x1) at (0, 1)  {$x_1$};          
\draw[norm] (x2) -- (x1);
\draw[norm] (x3) -- (x2);
\draw[norm] (x5) -- (x3);
\draw[norm] (x4) -- (x2);
\draw[norm] (x6) -- (x4);
\end{tikzpicture}
\vspace{1em}
\subcaption{Dynkin diagram shaped quiver in ${\rm Gr}(3,7)$.}
\label{quiver1}
\end{subfigure}
\hspace{2em}
\begin{subfigure}{0.45\textwidth}
\centering
\begin{tikzpicture}[scale=1.3]
\node[] (x3) at (-1, 0) {$x_3$};
\node[] (x5) at (-2, 0) {$x_5$}; 
\node[] (x2) at (0, 0)  {$x_2$};         
\node[] (x4) at (1, 0)  {$x_4$};        
\node[] (x6) at (2, 0)  {$x_6$};          
\node[] (x1) at (0, 1)  {$x_1$};          
\draw[norm] (x1) -- (x2);
\draw[norm] (x3) -- (x2);
\draw[norm] (x5) -- (x3);
\draw[norm] (x4) -- (x2);
\draw[norm] (x6) -- (x4);
\end{tikzpicture}
\vspace{1em}
\subcaption{Tree shaped quiver.}
\vspace{2em}
\label{quiver2}
\end{subfigure}
\begin{subfigure}{0.45\textwidth}
\centering
\begin{tikzpicture}[scale=1.3]
\node[] (x3) at (-1, 0) {$x_3$};
\node[] (x5) at (-2, 0) {$x_5$}; 
\node[] (x2) at (0, 0)  {$x_2$};         
\node[] (x4) at (1, 0)  {$x_4$};        
\node[] (x6) at (2, 0)  {$x_6$};          
\node[] (x1) at (0, 1)  {$x_1$};          
\draw[norm] (x2) -- (x1);
\draw[norm] (x3) -- (x2);
\draw[norm] (x5) -- (x3);
\draw[norm] (x2) -- (x4);
\draw[norm] (x4) -- (x6);
\end{tikzpicture}
\vspace{1em}
\subcaption{Quiver with bifurcations.}
\label{quiver3}
\end{subfigure}

\end{center}
\vspace{1em}
\caption{Examples of cluster quivers.}
\end{figure}
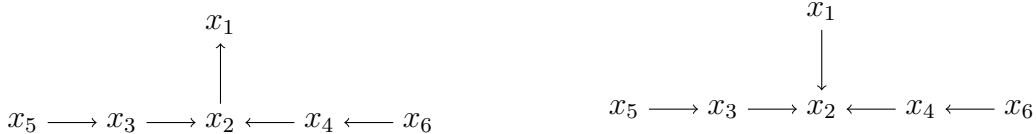
  
In \cite{Drummond:2018dfd} a method to systematically construct the face variables from a cluster quiver  diagram was described and given explicitly in the $E_6$ (or ${\rm Gr}(3,7)$) case.  First, one has to find a Dynkin diagram shaped quiver.  There are many quivers of this shape. In figure \eqref{quiver1} we show an example from the ${\rm Gr}(3,7)$ case.  If we denote by $x_i$ the $\mathcal{X}$-coordinates in node $i$ of the quiver, then the corresponding $u$-coordinates take the form
\begin{equation}\label{quiveru}
u_i = \frac{x_i f_i}{1 + x_i f_i},
\end{equation}
where
\begin{equation*}
\begin{aligned}
f_1=1,\quad f_2=1+x_1,\quad f_3=f_4=1+x_2(1+x_1), \\
f_5=1+x_3(1+x_2(1+x_1)), \quad f_6=1+x_4(1+x_2(1+x_1)).
\end{aligned}
\end{equation*}

In fact, we can generalise the method of \cite{Drummond:2018dfd} to include any tree shaped cluster.  Then, the $u$-coordinate of node $i$ can be found by following the path of the arrow that starts from node $i$ and follow the recursive formula $f_i=1+x_j f_j$, where $j$ is the first node we land by following the path of the arrow.  As an example, we have for figure \eqref{quiver2}
\begin{equation*}
f_1=f_3=f_4=1+x_2,\quad f_2=1,\quad f_5=1+x_3(1+x_2), \quad f_6=1+x_4(1+x_2).
\end{equation*}
When there is a bifurcation we consider the product of paths.  For example, for figure \eqref{quiver3} we have
\begin{equation*}
\begin{aligned}
f_1=f_6=1, \quad f_2=(1+x_1)(1+x_4(1+x_6)),\quad f_4= 1+x_6,\\
f_3=1+x_2(1+x_1)(1+x_4(1+x_6)), \quad f_5=1+x_3(1+x_2(1+x_1)(1+x_4(1+x_6))).
\end{aligned}
\end{equation*}

The method is valid when the quiver contains loops, if the chosen node does not contain any path that forms a loop.

We would now like to outline a different method for finding the face variables which is more directly related to the tropical fans and their associated scattering equations discussed in the preceding sections. Let us recall that the general form for the potential function is
\be
F = \sum_a s_a \log a\,,
\label{aF}
\ee
where we have used a very compact notation, with the sum being  over all the $\mathcal{A}$-coordinates $a$ of the cluster algebra (including the frozen ones) and $s_a$ the corresponding generalised Mandelstam variable. For the minors $\langle i_1 \ldots i_k \rangle$ the associated Mandelstams are the $s_{i_1 \ldots i_k}$ but the $s_a$ also include the Mandelstams associated to e.g. the quadratic $\mathcal{A}$-coordinates (\ref{quadraticAs}). 

We claim that the potential can also be written
\be
F = \sum_a v_a \log u_a\,,
\label{uF}
\ee
where the sum is over only the \emph{unfrozen} $\mathcal{A}$-coordinates $a$, $v_a$ is the ray (evaluated in terms of Mandelstam variables) and $u_a$ is the corresponding face variable. Note that, due to the generalised momentum conservation relations, (\ref{aF}) is homogeneous, even if each term individually is not. The expression (\ref{uF}) is manifestly homogeneous since the $u_a$ are homogeneous combinations of $\mathcal{A}$-coordinates. Once one has solved the tropical problem and found the rays, equating (\ref{aF}) and (\ref{uF}) gives a simple linear system to solve for the $\log u_a$ in terms of the $\log a$. The solution is exactly the face variables. Thus the tropical geometry provides a simple map from $\mathcal{A}$-coordinates to face variables. The method described above is closely related to the discussion of face variables in \cite{Arkani-Hamed:2019mrd} based on Minkowski sums of Newton polytopes arising from considering generalisations of string worldsheet integrals to ${\rm Gr}(k,n)$. 

\subsection{${\rm Gr}(3,6)$}
As described above, the ${\rm Gr}(3,6)$ cluster fan consists of 16 rays which are divided into four cyclic classes of size 6,6,2 and 2.  Written in terms of generalised Mandelstam invariants, the generators of these four classes are
\begin{equation}\label{36rays}
\{v_a\} = \{
s_{123}, t_{1234}+s_{q_1}, r_{341256}+s_{q_1}, r_{123456}+s_{q_1}
\}.
\end{equation}

For each of the 16 rays $v_a$ we can associate the $\mathcal{A}$-coordinate $a$, which can be found from mutations as described in \cite{Drummond:2019qjk}.  For the rays in \eqref{36rays} we associate
\begin{equation}\label{36Acoords}
\{
\left<124\right>,
\left<125\right>,
\left<135\right>,
\left<12[34]56\right>
\},
\end{equation}
where we recall $\left<12[34]56\right> \equiv \left<124\right>\left<356\right>-\left<123\right>\left<456\right>$.  

In addition, to each ray $v_a$ we can associate the face variable $u_a$.  As described above we may derive them from the equality of the two ways of writing the potential $F$ in (\ref{potentialGen}),
\begin{equation}\label{36findUcoords}
\sum_{i<j<k} s_{ijk}\log \langle ijk \rangle + \sum_{i=1}^2 s_{q_i} \log q_i = 
\sum_a v_a \log u_a\,.
\end{equation} 
All 22 $\mathcal{A}$-coordinates  appear in \eqref{36findUcoords}, including the frozen ones, however the generalised momentum conservation relations (\ref{momentum_conservation}) imply the LHS can be written as a combination of 16 homogeneous combinations of $\mathcal{A}$-coordinates. Equation \eqref{36findUcoords} therefore reduces to a linear system for of the 16 unknowns $\log u_a$ with a unique solution.  For the rays in \eqref{36rays} we find the corresponding $u$-coordinates,
\begin{equation}\label{36Ucoords}
\bigg\{
\frac{\left<123\right> \left<246\right>}{\left<124\right> \left<236\right>},
\frac{\left<12[34]56\right>}{\left<125\right>\left<346\right>},
\frac{\left<125\right>\left<134\right>\left<356\right>}{\left<135\right>\left<12[34]56\right>},
\frac{\left<124\right>\left<256\right>\left<346\right>}{\left<246\right>\left<12[34]56\right>}
\bigg\}\,,
\end{equation}
in agreement with the restriction of the $E_6$ $u$-coordinates found in \cite{Drummond:2018dfd} to $D_4$ and also in agreement with the $u$-coordinates given in \cite{Arkani-Hamed:2019mrd}.
We observe that the $\mathcal{A}$-coordinates appear in the denominators of the corresponding $u$-coordinates.

Labelling the 16 $u$-coordinates generated by the cyclic classes of \eqref{36Ucoords} as $\{u_1,\ldots,u_{6}\}$, $\{u_7,\ldots,u_{12}\}$, $\{u_{13},u_{14}\}$ and $\{u_{15},u_{16}\}$, we find that they satisfy the identities
\begin{align}
1
&=u_1+u_2 u_6 u_8 u_{11} u_{13} u_{16} \notag \\
&=u_7+u_3 u_6 u_8 u_{12} u_{14} u_{16} \notag \\
&=u_{13}+u_1 u_3 u_5 u_8 u_{10} u_{12} u_{14}^2 u_{15} u_{16} \notag \\
&=u_{15}+u_{2} u_{4} u_{6} u_{8} u_{10} u_{12} u_{13} u_{14} u_{16}^2\,,
\end{align}
which respect the boundary structure of the cluster polytope.

\subsection{${\rm Gr}(3,7)$}
The ${\rm Gr}(3,7)$ cluster fan posesses 42 rays, divided into 6 cyclic classes, each of size 7.  Written in terms of Mandelstam variables introduced in (\ref{gr37genmand}) the classes are generated by
\begin{align}
&s_{123}, \notag \\
&t_{1234} + s_{q_{55}}+s_{q_{66}}+s_{q_{57}}, \notag \\
&t_{1234567} +s_{q_{53}} + s_{q_{64}} + s_{q_{55}}+s_{q_{66}}+s_{q_{57}},\notag \\
&t_{1234567}+s_{134}+s_{234}+s_{q_{53}} + s_{q_{64}} + s_{q_{55}}+s_{q_{66}}+s_{q_{57}},\notag \\
&t_{1234567}+s_{167}+s_{267}+s_{q_{53}} + s_{q_{64}} + s_{q_{55}}+s_{q_{66}}+s_{q_{57}},\notag \\
&t_{1234}+t_{1267}+s_{125}+s_{q_{53}} + s_{q_{64}} + 2 s_{q_{55}}+s_{q_{66}}+s_{q_{57}}\,,
\end{align}
where $t_{1234567}=s_{123}+s_{124}+s_{125}+s_{126}+s_{127}$. Setting the $s_{q_{5i}}$ and $s_{q_{6i}}$ to zero in the above we recover the form of the rays given in \cite{Cachazo:2019ngv,Drummond:2019qjk}.
The corresponding $\mathcal{A}$-coordinates are
\begin{equation}\label{37Acoords}
\{
\left< 124 \right>,
\left< 125 \right>,
\left< 134 \right>,
\left< 135 \right>,
\left< 12[34]67 \right>,
\left< 12[35]67 \right>
\}.
\end{equation}

The equality of the two forms of the potential  (\ref{aF}) and (\ref{uF}) becomes
\begin{align}\label{37findUcoords}
&\sum_{i<j<k} s_{ijk} \log \langle ijk \rangle + \bigl[(s_{q_{51}} \log \langle 23 [45] 67 \rangle + s_{q_{61}} \log \langle 56[72]34\rangle) +{\text{cyc.}}\bigr] \notag \\
&= \sum_a v_a \log u_a\,.
\end{align}
Due to the generalised momentum conservation relation (\ref{gr37momcons}) both sides are homogeneous and we obtain a linear system for the $\log u_a$. They are found to be
\begin{align}
\label{37Ucoords}
\biggl\{
&\frac{\left< 123 \right>\left< 247 \right>}{\left< 124 \right>\left< 237 \right>},
\frac{\left< 12[34]57 \right>}{\left< 125 \right>\left< 347 \right>},
\frac{\left< 12[34]67 \right>}{\left< 134 \right>\left< 267 \right>}, 
\frac{\left< 134 \right> \left< 12[35]67 \right>}{\left< 135 \right> \left< 12[34]67 \right>},\notag \\
&\frac{\left< 267 \right> \left< 12[34]57 \right>}{\left< 257 \right> \left< 12[34]67 \right>},
\frac{\left< 125 \right>\left< 357 \right> \left< 12[34]67 \right>}
     {\left< 12[35]67 \right> \left< 12[34]57 \right>}
\biggr\}\,,
\end{align}
in agreement with the $u$-coordinates found in \cite{Drummond:2018dfd}. 
As in the ${\rm Gr}(3,6)$ case, the $\mathcal{A}$-coordinates appear in the denominators of the corresponding $u$-coordinates.

The 42 $u$-coordinates of \eqref{37Ucoords} obey the cluster connectivity and satisfy the identities
\begin{equation}
\begin{aligned}
1
&=u_1+u_2 u_7 u_9 u_{13} u_{17} u_{21} u_{22} u_{24} u_{27} u_{30} u_{32} u_{35} u_{36} u_{37} u_{39} u_{41}\\
&=u_8+ u_3 u_7 u_9 u_{10} u_{13} u_{14} u_{18} u_{21} u_{23} u_{25} u_{27} u_{28} u_{30} u_{32} u_{33} u_{35} u_{37}^2 u_{39} u_{40} u_{41} u_{42}\\
&=u_{15}+u_2 u_6 u_9 u_{12} u_{16} u_{21} u_{23} u_{26} u_{28} u_{30} u_{32} u_{35} u_{37} u_{39} u_{40} u_{42}\\
&=u_{22}+u_1 u_3 u_6 u_9 u_{10} u_{12} u_{14} u_{16} u_{18} u_{21} u_{23}^2 u_{25} u_{26} u_{28}^2 u_{29} u_{30} u_{31} u_{32} u_{33} u_{35} u_{37}^2 u_{38} u_{39} u_{40}^2 u_{42}^2\\
&=u_{29}+u_2 u_5 u_7 u_9 u_{11} u_{12} u_{14} u_{16} u_{19} u_{21} u_{22} u_{23} u_{25} u_{26} u_{27} u_{28} u_{30}^2 u_{32} u_{33} u_{35}^2 u_{37}^2 u_{39}^2 u_{40} u_{41} u_{42}^2\\
&=u_{36}+u_1 u_3 u_5 u_7 u_9 u_{10} u_{11} u_{12} u_{14}^2 u_{16} u_{18} u_{19} u_{21} u_{23}^2 u_{25}^2 u_{26} u_{27} u_{28}^2 u_{30}^2 u_{31} u_{32} u_{33}^2 u_{35}^2 u_{37}^3 u_{38} \\ &\qquad \qquad \times u_{39}^2 u_{40}^2 u_{41} u_{42}^3.
\end{aligned}
\end{equation}
Powers of 3 appear for the first time.

\subsection{${\rm Gr}(3,8)$}
The ${\rm Gr}(3,8)$ cluster fan consists of 128 rays divided into 16 cyclic classes of size 8.  Explicitly, in terms of the $\tilde{x}$ variables, the 128 rays (or $g$-vectors) are 
\begin{equation}\label{38gvectors}
\begin{aligned}
g_1&=(1, 0, 0, 0, 0, 0, 0, 0), & g_9&=(0, 0, 1, 0, 0, 0, 0, 0),\\
g_{17}&=(0, 0, 0, 0, 1, 0, 0, 0),& g_{25}&=(0, 1, 0, 0, 0, 0, 0, 0),\\
g_{33}&=(-1, 0, 0, 1, 0, 0, 0, 0),& g_{41}&=(0, 0, -1, 0, 0, 1, 0, 0), \\
g_{49}&=(0, 1, 1, -1, 0, 0, 0, 0),&  g_{57}&=(0, 1, 1, 0, 0, -1, 0, 0),\\
g_{65}&=(0, 1, 1, 0, 0, 0, 0, -1),&g_{73}&= (0, 1, 0, 0, 1, -1, 0, 0), \\
g_{81}&=(0, 1, 0, 0, 1, 0, 0, -1), &g_{89}&=(-1, 0, 0, 1, 1, -1, 0, 0), \\
g_{97}&=(-1, 0, 0, 1, 1, 0, 0, -1), & g_{105}&=(0, 1, 0, 1, 1, -1, 0, -1),\\
g_{113}&=(-1, 0, 0, 2, 1, -1, 0, -1), & g_{121}&=(-1, 1, 0, 1, 1, -1, 0, -1)
\end{aligned}
\end{equation}
and their cyclic rotations in the Mandelstam space.  We recall that it is straightforward to map any $g$-vector to the Mandelstam space by evaluating all the $\mathcal{A}$-coordinates as tropical polynomials.   
The list of all 128 vertices in the Mandelstam space corresponding to the Speyer-Williams fan was given in \cite{Drummond:2019qjk}. In fact, if we only include the $s_{ijk}$ Mandelstam variables, eight of the vectors are redundant in that they are not true vertices of the fan. By extending the kinematics to include also 80 generalised Mandelstam variables corresponding to the 56 quadratic and 24 cubic $\mathcal{A}$-coordinates, we obtain the 128 rays of the cluster fan with no redundancies. The expressions are cumbersome so we omit them here.

The $\mathcal{A}$-coordinates are generated by cyclic rotations of the following,
\begin{align}
\label{38Acoords}
\{
&\left< 124 \right>,
\left< 125 \right>,
\left< 126 \right>,
\left< 134 \right>,
\left< 135 \right>,
\left< 136 \right>,\notag \\
&\left< 12[34]56 \right>,
\left< 12[34]57 \right>,
\left< 12[34]58 \right>,
\left< 12[34]67 \right>,
\left< 12[34]68 \right>,
\left< 12[35]67 \right>,
\left< 12[35]68 \right>,\notag \\
&\left< 12[34]8[67]45 \right>,
\left< 12[35]8[67]45 \right>,
\left< 12[34]8[67]35 \right>
\},
\end{align}
where in the final line we have defined the cubic coordinates via
\begin{equation}
\begin{aligned}
\left< 12[34]5[67]89 \right> &\equiv 
\left< 124 \right> \left<35[67]89\right> - \left< 123 \right> \left<45[67]89\right> \\
&= \left< 12[34]57 \right> \left<689 \right> - \left< 12[34]56 \right> \left<789 \right> 
=-\left< 67[89]5[12]34 \right>.
\end{aligned}
\end{equation}
 
In each of the 24 cubic $\mathcal{A}$-coordinates 7 indices appear once and 1 index appears twice.  Denoting the quadratic $\mathcal{A}$-coordinates by $q$ and the cubic ones by $c$, the generalised momentum conservation reads schematically
\be
\sum_{j<k} s_{ijk} + \sum_{q|i\in q} s_q + \sum_{c | i \in c} s_c + 2 \sum_{c | i^2 \in c} s_c = 0\,,  \qquad i=1,\ldots,8\,,
\ee
where the factor of two in the final term accounts for the double appearance of $i$ in the associated cubic coordinate. 

As above we determine the $u$-coordinates from the equality of the two forms of the potential,
\begin{equation}
\sum_{i<j<k} s_{ijk}\log \langle ijk \rangle + \sum_q s_q\, \log q + \sum_c s_c \, \log c = 
\sum_a v_a \log u_a\,.
\end{equation}
The $u$-coordinates are found to be
\begin{equation}
\begin{aligned}
\biggl\{
&\frac{\langle 123 \rangle \langle 248 \rangle}
{\langle 124 \rangle \langle 238 \rangle},
\frac{\langle 12[34]58 \rangle}{\langle 125 \rangle \langle 348 \rangle},%
\frac{\langle 12[45]68 \rangle}{\langle 126 \rangle \langle 458 \rangle},%
\frac{\langle 12[34]78 \rangle}{\langle 134 \rangle \langle 278 \rangle},%
\frac{\langle 134 \rangle \langle 12[35]78 \rangle}{\langle 135 \rangle \langle 12[34]78 \rangle},
\frac{\langle 13[45]6[78]12 \rangle}{\langle 136 \rangle \langle 12[45]78 \rangle},\\%
&\frac{\langle 124 \rangle \langle 34[56]28 \rangle}{\langle 248 \rangle \langle 12[56]34 \rangle},
\frac{\langle 12[56]34 \rangle \langle 34[57]28 \rangle}
{\langle 12[57]34 \rangle \langle 34[56]28 \rangle},
\frac{\langle 348 \rangle \langle 12[34]5[67]82 \rangle}
{\langle 12[58]34 \rangle \langle 34[67]28 \rangle},
\frac{\langle 12[34]5[67]82 \rangle}{\langle 258 \rangle \langle 12[34]67 \rangle},\\%
&\frac{\langle 268 \rangle \langle 12[34]8[67]45 \rangle}{\langle 12[68]34 \rangle \langle 45[67]28 \rangle},
\frac{\langle 125 \rangle \langle 12[34]8[35]67 \rangle}{\langle 12[34]58 \rangle \langle 12[67]35 \rangle},
\frac{\langle 12[68]34 \rangle \langle 12[35]8[67]45 \rangle}{\langle 12[68]35 \rangle \langle 12[34]8[67]45 \rangle},\\
&\frac{\langle 45[67]28\rangle \langle 12[34]8[67]35 \rangle}{\langle 35[67]28 \rangle \langle 12[34]8[67]45 \rangle},
\frac{\langle 358 \rangle \langle 12[35]67 \rangle \langle 12[34]8[67]45 \rangle}{\langle 12[34]8[67]35 \rangle \langle 12[35]8[67]45 \rangle},\\%
&\frac{\langle 12[34]67 \rangle \langle 35[67]82 \rangle \langle 12[34]58 \rangle}{\langle 12[34]5[67]82 \rangle \langle 12[34]8[67]35 \rangle}%
\biggr\}
\end{aligned}
\end{equation}
and satisfy identities reflecting the cluster connectivity.  The highest power appearing in the identities is 6.  

In the generalised $\phi^3$ amplitude corresponding to the Speyer-Williams fan, the last 8 $g$-vectors in \eqref{38gvectors} correspond to spurious poles. The facets containing them always combine together into bigger facets without them. When we introduce the generalised Mandelstam variables for each $\mathcal{A}$-coordinate (not just the minors) then this is no longer the case. The fan is simplicial and every vertex contributes on an equal footing. In the interpretation where the volume of each facet is thought of as a generalised Feynman diagram, each diagram now has the same number of poles.

It is possible to define $u$-coordinates for the  polytope dual to the Speyer-Williams fan formed from only the first 120 $g$-vectors.  Only 16 of the $u$-coordinates are affected.  These are $u^{(120)}_{57} = u^{(128)}_{57} u^{(128)}_{121}$ and $u^{(120)}_{65} = u^{(128)}_{65} u^{(128)}_{125}$ and their cyclic rotations.  Then the corresponding $u$-identities obey the connectivity of the resulting polytope after the removal of the last 8 $g$-vectors.

\section{Tropically adjacent polylogarithms}
\label{polylogs}

The cases of the Grassmannians ${\rm Gr}(4,n)$ are particularly interesting because they describe the kinematic space associated to scattering amplitudes in planar $\mathcal{N}=4$ super Yang-Mills theory. This is due to the cyclically ordered kinematics being neatly parametrised in terms of momentum twistors, and dual conformal symmetry meaning that amplitudes depend on (homogeneous combinations of) the Pl\"ucker coordinates $\langle ijkl \rangle$. In \cite{Golden:2013xva} a connection was observed between the logarithmic branch cuts of loop amplitudes and the $\mathcal{A}$-coordinates of the ${\rm Gr}(4,n)$ cluster algebras. Specifically the symbol alphabet of the two-loop MHV amplitudes found in \cite{CaronHuot:2011ky} is comprised of $\mathcal{A}$-coordinates.

The connection between cluster algebras and scattering amplitudes has been useful in constructing amplitudes via the analytic bootstrap of e.g. \cite{Dixon:2011pw,Dixon:2013eka,Dixon:2014voa}. In particular, it was important for identifying the heptagon symbol alphabet used to construct the three-loop and four-loop results for seven-point amplitudes \cite{Drummond:2014ffa,Dixon:2016nkn,Drummond:2018caf}. In \cite{Drummond:2017ssj} a further connection to the cluster structure was made. The adjacent pairs of letters in the symbols of hexagon and heptagon amplitudes are constrained to be ones which appear together in the same cluster. Geometrically this means that they are the $\mathcal{A}$-coordinates corresponding to codimension one boundaries of the cluster polytope which intersect (or frozen coordinates which appear in every cluster). These cluster adjacency constraints are consistent with the Steinmann relations on overlapping discontinuities \cite{Bartels:2008sc,Caron-Huot:2016owq}.

\newcommand{\bdiamond}[1][fill=black]{\tikz [x=1.2ex,y=1.2ex,line width=.1ex,line join=round, yshift=-0.285ex] \draw  [#1]  (0,0) -- (.5,.5) -- (0,1) -- (-0.5,0.5)  -- cycle;}%
\newcommand{\wdiamond}[1][fill=white]{\tikz [x=1.2ex,y=1.2ex,line width=.1ex,line join=round, yshift=-0.285ex] \draw  [#1]  (0,0) -- (.5,.5) -- (0,1) -- (-0.5,0.5)  -- cycle;}%
\newcommand{\bcircle}[1][fill=black]{\tikz [x=1.2ex,y=1.2ex,line width=.1ex,line join=round, yshift=-0.285ex] \draw  [#1]  (0,0) circle (0.5);}%
\newcommand{\wcircle}[1][fill=white]{\tikz [x=1.2ex,y=1.2ex,line width=.1ex,line join=round, yshift=-0.285ex] \draw  [#1]  (0,0) circle (0.5);}%

\begin{table*}
  \centering

  {{
  \setlength\tabcolsep{1.6pt}
\scalebox{1}{\begin{tabular}{r|ccccccc|ccccccc|ccccccc|ccccccc|ccccccc|ccccccc}
& \multicolumn{7}{|c}{\as{1}{i}} & \multicolumn{7}{|c}{\as{2}{i}} & \multicolumn{7}{|c}{\as{3}{i}} & \multicolumn{7}{|c}{\as{4}{i}} & \multicolumn{7}{|c}{\as{5}{i}} & \multicolumn{7}{|c}{\as{6}{i}} \\
\hline
\as{1}{1}& $\bcircle$& $\wcircle$& $\wcircle$& $\bdiamond$& $\bdiamond$& $\wcircle$& $\wcircle$& $\bdiamond$& $\bdiamond$& $\wcircle$& $\bcircle$& $\bdiamond$& $\bcircle$& $\wcircle$& $\bdiamond$& $\wcircle$& $\bcircle$& $\bdiamond$& $\bcircle$& $\wcircle$& $\bdiamond$& $\bcircle$& $\wcircle$& $\bdiamond$& $\wcircle$& $\wcircle$& $\bdiamond$& $\wcircle$& $\bcircle$& $\wcircle$& $\bdiamond$& $\wcircle$& $\wcircle$& $\bdiamond$& $\wcircle$& $\wdiamond$& $\bdiamond$& $\wcircle$& $\wcircle$&  $\wcircle$& $\wcircle$& $\bdiamond$\\
\as{2}{1}& $\bdiamond$& $\wcircle$& $\bcircle$& $\bdiamond$& $\bcircle$& $\wcircle$& $\bdiamond$& $\bcircle$& $\wcircle$& $\bcircle$& $\bdiamond$& $\bdiamond$& $\bcircle$&   $\wcircle$& $\bdiamond$& $\wcircle$& $\bdiamond$& $\bdiamond$& $\wcircle$& $\bdiamond$& $\bcircle$& $\bdiamond$& $\wcircle$& $\bdiamond$& $\wcircle$& $\bcircle$& $\bdiamond$& $\wcircle$& $\wcircle$& $\bdiamond$& $\bcircle$& $\wcircle$& $\bdiamond$& $\wcircle$& $\bdiamond$& $\wcircle$& $\bdiamond$& $\wcircle$& $\bcircle$& 
  $\wcircle$& $\bdiamond$& $\wcircle$\\
\as{3}{1}& $\bdiamond$& $\bdiamond$& $\wcircle$& $\bcircle$& $\bdiamond$& $\bcircle$& $\wcircle$& $\bdiamond$& $\bcircle$& $\bdiamond$& $\wcircle$& $\bdiamond$& $\bdiamond$&  $\wcircle$& $\bcircle$& $\wcircle$& $\bcircle$& $\bdiamond$& $\bdiamond$& $\bcircle$& $\wcircle$& $\bdiamond$& $\wcircle$& $\bdiamond$& $\bcircle$& $\wcircle$& $\bdiamond$& $\wcircle$& $\wcircle$& $\bdiamond$& $\wcircle$& $\bdiamond$& $\wcircle$& $\bcircle$& $\bdiamond$& $\wcircle$& $\wcircle$& $\bdiamond$& $\wcircle$& 
  $\bcircle$& $\wcircle$& $\bdiamond$\\
\as{4}{1}&$\bcircle$& $\wcircle$& $\bdiamond$& $\wcircle$& $\wcircle$& $\bdiamond$& $\wcircle$& $\bdiamond$& $\wcircle$& $\bdiamond$& $\bcircle$& $\wcircle$& $\bdiamond$& $\wcircle$& $\bdiamond$& $\wcircle$& $\bdiamond$& $\wcircle$& $\bcircle$& $\bdiamond$& $\wcircle$& $\bcircle$& $\wdiamond$& $\bdiamond$& $\wcircle$& $\wcircle$& $\bdiamond$& $\wdiamond$& $\bcircle$& $\wcircle$& $\wcircle$& $\wcircle$& $\wcircle$& $\wcircle$& $\wcircle$& $\wdiamond$& $\bdiamond$& $\wdiamond$& $\wcircle$& 
  $\wcircle$& $\wdiamond$& $\bdiamond$\\
\as{5}{1}& $\bcircle$& $\wcircle$& $\bdiamond$& $\wcircle$& $\wcircle$& $\bdiamond$& $\wcircle$& $\wcircle$& $\bdiamond$& $\wcircle$& $\bdiamond$& $\wcircle$& $\bcircle$& $\bdiamond$& $\wcircle$& $\bdiamond$& $\bcircle$& $\wcircle$& $\bdiamond$& $\wcircle$& $\bdiamond$& $\bcircle$& $\wcircle$& $\wcircle$& $\wcircle$& $\wcircle$& $\wcircle$& $\wcircle$& $\bcircle$& $\wdiamond$& $\bdiamond$& $\wcircle$& $\wcircle$& $\bdiamond$& $\wdiamond$& $\wdiamond$& $\bdiamond$& $\wdiamond$& $\wcircle$& $\wcircle$& $\wdiamond$& $\bdiamond$\\
\as{6}{1}& $\wdiamond$& $\bdiamond$& $\wcircle$& $\wcircle$& $\wcircle$& $\wcircle$& $\bdiamond$& $\wcircle$& $\wcircle$& $\bdiamond$& $\wcircle$& $\bcircle$& $\wcircle$& $\bdiamond$& $\wcircle$& $\bdiamond$& $\wcircle$& $\bcircle$& $\wcircle$& $\bdiamond$& $\wcircle$& $\wdiamond$& $\bdiamond$& $\wdiamond$& $\wcircle$& $\wcircle$& $\wdiamond$& $\bdiamond$& $\wdiamond$& $\bdiamond$& $\wdiamond$& $\wcircle$& $\wcircle$& $\wdiamond$& $\bdiamond$& $\bcircle$& $\wdiamond$& $\wcircle$& $\wdiamond$& $\wdiamond$& $\wcircle$& $\wdiamond$\\
\end{tabular}
}
}}
\caption{\small The neighbourhood and connectivity relations of the coordinates \as{i}{1} with the 42-letter alphabet. Other relations can be inferred by cyclic symmetry. \\
$\bdiamond$: There are clusters where the coordinates appear together connected by an arrow.\,\,\\
$\bcircle$: There are clusters where the coordinates appear together but they are never connected.\,\, \\
$\wcircle$: The coordinates never both appear in a cluster but there is a mutation that links them.\,\,\\
$\wdiamond$: The coordinates never both appear in a cluster nor is there a mutation that links them. \\
}
  \label{fig:adjacency}  
\end{table*}

The notion of cluster adjacency gives rise to an interesting class of polylogarithmic functions, associated to a given cluster algebra. Here we would like to generalise this notion to different possible choices of fan $F(\mathcal{S})$. The cluster adjacent polylogs in the sense of \cite{Drummond:2017ssj} will correspond to the cluster fans. Those corresponding to less refined fans will obey additional constraints beyond the fact that adjacent pairs of letters must appear together in a cluster.

In Section \ref{Sect-fans} we discussed different tropical fans related to ${\rm Gr}(3,7) \cong {\rm Gr}(4,7)$, generated by different sets of tropical $\cA$-coordinates. Here we will discuss this further and the implications these different fans have for cluster adjacency and scattering amplitudes in SYM.

\subsection*{Edges}

If two rays are connected by an edge in the cluster fan this means their corresponding $\cA$-coordinates appear together in a cluster in the cluster algebra and hence are cluster adjacent. The $\{a_1,\ldots,a_6\}$ fan in Table \ref{Gr37fans} is the most refined fan one could construct and is dual to the Gr$(4,7)$ cluster polytope. This fan consists of 399 edges which correspond to all of the cluster adjacent pairs of different $\cA$-coordinates. The $\{a_1,\ldots, a_5\}$ fan also contains all 399 edges and so offers no alteration to cluster adjacency at the level of edges. However, we will see that this fan does differ from the cluster fan at the level of triangles. The $\{a_1,\ldots, a_4\}$ fan also called the Speyer-Williams (SW) fan was the original tropical fan for Gr$(3,7)$ discussed in \cite{2003math.....12297S}. This fan has 392 edges and is the first instance where we have a differing number of edges from that of the cluster fan as this fan has 392 edges. The $\{a_1,\dots, a_3\}$ fan has 385 edges, 14 fewer than the cluster fan. These 14 edges correspond to the pairs 
\begin{equation}
\{a_{21}, a_{64}\} \quad + \quad \text{dihedral}
\label{MHVmissingpairs}
\end{equation}
which are the pairs observed to be missing from certain integrals and MHV amplitudes in \cite{Drummond:2017ssj}. We note that the missing pairs are neighbours of `disconnected' type in the language of \cite{Drummond:2017ssj}. This can be seen in Table \ref{fig:adjacency} (which is the adjacency table of \cite{Drummond:2017ssj} reproduced here) in the row labelled by $a_{21}$ and the fourth column in the $a_{6i}$ block where the symbol $\bcircle$ appears. That is, they appear together in the same cluster, but never connected by an arrow. As noted in \cite{Drummond:2017ssj}, this has the consequence that, if the pair were to appear consecutively in a symbol, the integrability conditions would impose that they do so in a symmetric way. In other words, the corresponding weight-two function is simply a product of logarithms, $\log a_{21} \log a_{64}$. Therefore there is no distinction between the ordering shown in (\ref{MHVmissingpairs}) and the reverse.

\subsection*{Triangles}

Much like with edges mentioned above, if three rays are all connected to each other to form a triangle in the cluster fan then all three corresponding $\cA$-coordinates can be found in a cluster together and hence are cluster adjacent. When considering the less refined fans, if we find that an edge is missing it follows that any triangles containing that edge are also missing. However, it is also possible for further triangles to be absent, even if all three edges of the triangle are still present. We have seen this phenomenon in the ${\rm Gr}(3,6)$ example discussed in Sect. \ref{Sect-fans}. When any pair of connected tetrahedra in Fig. \ref{D4clusterfanpoly} becomes a bipyramid, the triangle at the interface is removed. The same phenomenon can happen in the ${\rm Gr}(3,7) \cong {\rm Gr}(4,7)$ case.

The cluster fan contains 1547 triangles but the $\{a_1,\ldots, a_5\}$ fan only has 1540 triangles, 7 fewer than the cluster fan. These missing triangles are
\begin{equation}
\{a_{11}, a_{41}, a_{51}\} \quad + \quad \text{cyclic.}
\label{nmhvmissingtriangles}
\end{equation}
As mentioned above the $\{a_1,\ldots, a_3\}$ fan has 14 fewer edges than the cluster fan. These edges appear in $(8 \times 14) + 7 = 119$ triangles, corresponding to 8 triangles and their dihedral copies along with 7 triangles which contain 2 of the 14 missing edges. The other missing triangles are 
\begin{align}
\label{mhvmissingtriangles1}
\{a_{22}, a_{24}, a_{16}\} \quad &+ \quad \text{dihedral}\,,\\
\{a_{21}, a_{13}, a_{53}\} \quad &+ \quad \text{dihedral} \quad + \quad \text{parity}\,.
\label{mhvmissingtriangles2}
\end{align}

The triangles (\ref{nmhvmissingtriangles}), (\ref{mhvmissingtriangles1}) and (\ref{mhvmissingtriangles2}) are fully disconnected in the sense that all three edges correspond to disconnected neighbours. For example, we see in Table \ref{fig:adjacency} the symbol $\bcircle$ corresponding to the pairs $\{a_{11},a_{41}\}$, $\{a_{11},a_{51}\}$ and $\{a_{41},a_{51}\}$.  There are a total of 70 disconnected triangles in the cluster fan, 56 of which are missing from the $\{a_1, \ldots, a_3\}$ fan (the 49 in (\ref{nmhvmissingtriangles}), (\ref{mhvmissingtriangles1}) and (\ref{mhvmissingtriangles2}) and the 7 which contain two missing edges of the form (\ref{MHVmissingpairs})). The remaining 14 disconnected triangles are of the form
\begin{equation}
\{a_{11}, a_{24}, a_{33}\} \quad + \quad \text{dihedral}
\label{disctriangles}
\end{equation}
and these ones are \emph{present} in the $\{a_1, \ldots, a_3\}$ fan. 

\subsection*{Comparison to amplitudes}

The variation in the number of edges and triangles in the above fans is interesting in the context of $\mathcal{N}=4$ SYM loop amplitudes. At the level of edges all currently known MHV heptagon amplitudes are consistent with the edges from the $\{a_1, \ldots, a_3\}$ fan. The NMHV heptagon amplitude at four loops \cite{Drummond:2018caf} requires all 399 pairs (edges) \cite{Henke:2019hve} and so is consistent with either the $\{a_1, \ldots, a_5\}$ fan or the $\{a_1, \ldots, a_6\}$ fan.

We have also observed that the triangles missing from the $\{a_1, \ldots, a_3\}$ fan are also missing from all available MHV and NMHV amplitudes. Thus at the level of triangles there is no distinction between the currently known MHV and NMHV amplitudes, though, as we have stated above, there is at the level of edges. The disconnected triangles (\ref{disctriangles}) which are present in the $\{a_1, \ldots ,a_3\}$ fan \emph{do} appear as consecutive triples of letters in known MHV and NMHV amplitudes.

The cluster adjacency conditions in heptagon functions seem to follow from the extended Steinmann conditions \cite{Caron-Huot:2019bsq} and the physical initial entry condition (and integrability of the symbol), at least up to weight seven. It is interesting to note therefore that the conditions obtained by imposing the absence of the triples (\ref{nmhvmissingtriangles}), (\ref{mhvmissingtriangles1}) and (\ref{mhvmissingtriangles2}) do not follow only from physical initial entry conditions and cluster adjacency, there being examples of functions in weight seven which do have the missing triangles in their symbols. Therefore, forbidding above the triangles is an extra condition that goes beyond cluster adjacency.

In summary, the known seven-point MHV amplitudes in planar $\mathcal{N}=4$ SYM are consistent with the structure of the  $\{a_1 ,\ldots, a_3\}$ tropical fan although there is limited data to verify this. One potential test would be to bootstrap the five-loop, MHV heptagon using the restrictions following from the $\{a_1, \ldots, a_3\}$ and investigate whether a solution could be found. For NMHV seven-point amplitudes, the edge structure suggests that the minimal fan compatible with their singularities would be the $\{a_1, \ldots, a_5\}$ fan. This fan has all possible edges but seven missing triangles. It would be interesting to investigate whether such triangles indeed continue to be absent at higher orders.

Beyond seven points the ${\rm Gr}(4,n)$ cluster algebras become infinite. For $n=8$ this infinity is of affine type and tropical fans have been considered in this case in several recent papers \cite{Drummond:2019qjk,Drummond:2019cxm,Arkani-Hamed:2019rds,Henke:2019hve}. The affine nature of the cluster algebra leads to natural infinite sequences of cluster coordinates and ${\bf g}$-vectors which lead to the appearance of quantities involving square roots. In \cite{Drummond:2019cxm} we found a way to associate precisely the same set of square roots as those appearing in a recent calculation of the two-loop eight-point NMHV amplitude \cite{Zhang:2019vnm}. It would be interesting also in that case to understand how the structure of edges and triangles (and beyond) is related to the analytic behaviour of the amplitudes.

\subsubsection*{Acknowledgements}

CK would like to thank Nick Early for useful discussions.
JD, JF and CK are supported by ERC grant 648630
IQFT. This project has received funding from the European Research Council (ERC)
under the European Union's Horizon 2020 research and innovation programme (grant
agreement No. 724638).

\Urlmuskip=0mu plus 1mu\relax
\def\UrlBreaks{\do\/\do-}
\bibliographystyle{JHEPmod}
\bibliography{biblio}

\end{document}